\documentclass[amsmath,amsfonts,aps,superscriptaddress,notitlepage]{revtex4-1}
\usepackage{graphicx}
\usepackage{subfigure}
\usepackage{mathtools}
\usepackage{ulem}

\newcommand{\ie}{i.e.}

\DeclareTextFontCommand{\emph}{\it}

\usepackage[usenames,dvipsnames]{xcolor} 

\newcommand{\namemc}{Multiplex Markov chain}

\newcommand{\sstate}[2]{\ensuremath{{#1}{#2}}}
\newcommand{\slstate}[1]{\ensuremath{{#1}}} 

\begin{document}
\title{Quantifying dynamical spillover in co-evolving multiplex networks}
\author{Vikram S. Vijayaraghavan}
\affiliation{Complexity Sciences Center, University of California, Davis CA 95616, USA}
\affiliation{Department of Physics, University of California, Davis CA 95616, USA}
\author{Pierre-Andr{\'e} No{\"e}l}
\affiliation{Complexity Sciences Center, University of California, Davis CA 95616, USA}
\affiliation{Department of Computer Science, University of California, Davis CA 95616, USA}
\author{Zeev Maoz}
\affiliation{Department of Political Science, University of California, Davis CA 95616, USA}
\affiliation{Interdisciplinary Center, Herzliya, Israel}
\author{Raissa M. D'Souza}
\affiliation{Complexity Sciences Center, University of California, Davis CA 95616, USA}
\affiliation{Department of Computer Science, University of California, Davis CA 95616, USA}
\affiliation{Department of Mechanical and Aerospace Engineering, University of California, Davis CA 95616, USA}
\affiliation{The Santa Fe Institute, Santa Fe NM 87501, USA}

\begin{abstract}
Multiplex networks (a system of multiple networks that have different
types of links but share a common set of nodes) arise naturally in a
wide spectrum of fields. Theoretical studies show that in such
multiplex networks, correlated edge dynamics between the layers can
have a profound effect on dynamical processes. However, how to extract
the correlations from real-world systems is an outstanding
challenge. Here we provide a null model based on Markov chains to
quantify correlations in edge dynamics found in longitudinal data of
multiplex networks. We use this approach on two different data sets:
the network of trade and alliances between nation states, and the
email and co-commit networks between developers of open source
software. We establish the existence of ``dynamical spillover''
showing the correlated formation (or deletion) of edges of different
types as the system evolves. The details of the dynamics over time
provide insight into potential causal pathways.
\end{abstract}

\maketitle

\section{Introduction}
\begin{figure}
\begin{center}
{\includegraphics[width=\linewidth]{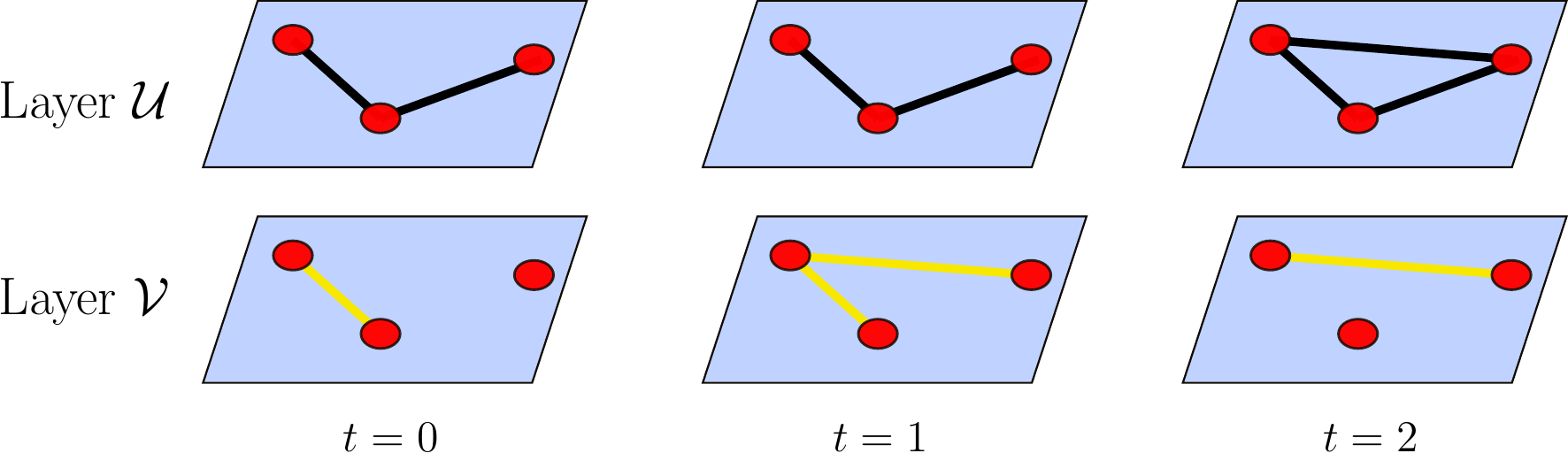}}
\caption{Three consecutive snapshots in time of an abstract multiplex
  network consisting of two layers, $\mathcal{U}$ and $\mathcal{V}$. A
  multiplex network is a network where different layers share the same
  set of nodes and edges on each layer represent a different
  relationship. For example, here the nodes could represent countries
  in the world, the edges in layer $\mathcal{U}$ the alliances between
  them, and the edges in layer V may denote trade between the same set
  of countries.
  \label{layered-net}}
\end{center}
\end{figure}

We are increasingly aware that no system lives in
isolation. Understanding the nature of a system of interacting
networks is a goal that researchers are increasingly focusing their
efforts on~\cite{nagurney2002, d2014, buldyrev2010, leicht2009}. One
category of such interacting networks are multiplex
networks. Multiplex networks are networks that share a common set of
nodes that can be linked via different types of edges that comprise
the system. Examples of such systems include networks of nation states
that have many types of relationships such as trade, alliances,
military conflicts etc.; communication networks of individuals who
might communicate by email, or over a social networking website; and a
transportation network of a city which consists of systems such as
road and railway connections. Such multiplex networks can be thought
of as layered networks that have the same set of nodes in each layer
and links of a particular type are restricted to a single layer (See
Fig.~\ref{layered-net}).

Multiplex networks, also called multirelational networks, were first
explored by sociologists as early as 1953~\cite{katz1953, boorman1976,
  hubert1978, pattison2000}. Multirelational networks were also
studied by computer scientists within the context of building better
community mining methods~\cite{cai2005, cai2005mining}. Recently, the
interest in modeling the dynamics and correlations present in
multiplex networks has increased dramatically. Yet, these works
explore the presence of correlations in static snapshots. Our focus is
on correlations in longitudinal evolution. A comprehensive survey of
studies on multiplex networks can be found in review articles by
Kivel\"a et al. and Boccaletti et
al~\cite{kivela2014,boccaletti2014}. Work by Nicosia et al. and Kim et
al. present models for the growth of multiplex networks and explore
the impact of metrics such as degree-distribution and degree-degree
correlations~\cite{nicosia2013, kim2013, nicosia2014}. Correlations
present in static snapshots of multiplex networks, such as
interdependence of node properties between different layers,
correlations in edge overlap, etc., have also been
explored~\cite{battiston2014, min2014, menichetti2014}. More recently,
Nicosia et al. extend the notion of degree-degree correlations in the
context of multiplex networks, measure these correlations in real
data, and present models to reproduce empirical
observations~\cite{nicosia2014measuring}.

From a co-evolution perspective, in many multiplex networks
representing real world complex systems, the edge dynamics on one
layer is often highly correlated with the edge dynamics on other
layers. A thorough understanding of these correlations could help us
design better network interventions for the control of these
systems. For example, data available on Open Source Software, can be
thought of as a multiplex network with two layers: first, the
communication between developers (such as email) and second, the
collaboration between developers (such as committing code to the same
file). Understanding correlations that exist between communications
and collaborations would help us plan social and/or coding-related
events to increase contribution more efficiently. Organizing social
events usually involve a lesser cost when compared to a coding event,
and help form friendships among developers that might then lead to
them collaborating on a project. Similarly, in other multiplex
networks, the cost of intervention may be less expensive on one layer
but might help in creating the desired effects on another layer where
it might be more expensive or harder to intervene.

In this article we present a systematic way, based on Markov
chains~\cite{norris1998markov}, to quantify correlations in edge
dynamics of multiplex networks using empirical data. As a first step
towards quantifying correlations in edge dynamics, we capture how the
presence (or absence) of a link between two nodes in one layer
influences the dynamics of the edge between the same two nodes in
another layer. Our work presents a quantitative way to uncover the
co-evolutionary dynamics of edges in different layers which may lead
to the emergence of overlap of edges among distinct layers. Such
overlaps have been shown to have a significant impact on dynamical
processes such as percolation occurring on the
network~\cite{cellai2013, li2013}.

We start with longitudinal data, \ie, time ordered snapshots, of the
multiplex network. From this data, we construct a Markov chain that
estimates the probability of observing an edge between a pair of nodes
on any layer of the multiplex at the current time step, based on the
presence/absence of the edge between the same pair of nodes, on all
layers, in the previous time step. We call this model the Multiplex
Markov chain, the details of which are explained in
Sec.~\ref{mc-method}. While the Multiplex Markov chain allows us to
empirically determine edge transition probabilities, we cannot
directly infer if co-occurring edge dynamics between layers are from
true correlations that exist between the layers of the multiplex
network or due to chance. In order to discern the presence of
correlations between layers we construct a null model that treats each
layer independently (\ie, one Markov chain for each layer ignoring the
presence of other layers). This enables us to compare the value of the
transition parameters obtained from the data with that of the null
model. We say that there is \textit{dynamical spillover} when the
transition parameters obtained from the Multiplex Markov chain are
considerably different from those obtained via the null model (see
methods section for details). We have made available the Python code
to construct the {\namemc} and the null model on
Github~\cite{mmc-code}.

Further, we illustrate the tool by applying it to two different
multiplex data sets: a network of trade and alliances between nation
states; and the co-commit and email networks of developers of an open
source software. We show that these networks exhibit dynamical
spillover, \ie, the edge dynamics on a layer of these multiplex
networks are strongly influenced by the presence of edges in another
layer. We also identify potential causal pathways that may influence
the co-evolution of these multiplex networks. Although we describe how
to construct the model for two layers of a multiplex network, the
{\namemc} and the null model straightforwardly generalize to multiplex
networks with more layers.

\section{Results}
\subsection{Multiplex Markov chain and the null model}
\label{mc-method}
In this section we present the {\namemc} model that quantifies the
correlations present in the edge dynamics occurring on two layers of a
multiplex network, and the corresponding null model that helps us
distinguish non-trivial correlations from those that occur due to
randomness. Inferring correlations in dynamics requires the use of
temporal data. We take a data-first approach and begin with time
ordered snapshots of the multiplex networks. For simplicity let us
focus on two layers of the multiplex network, denoted by calligraphic
letters (e.g., $\mathcal{U}$ and $\mathcal{V}$). The two layers of the
network share the same set of nodes between them but each layer only
has edges of a particular type. There are no links connecting
instances of nodes that are on different layers, \ie, links are
restricted to lie on a single layer. Each pair of nodes in this
multiplex can be represented using an ordered pair indicating the
presence or absence of a link in each layer. We use lowercase Latin
letters (e.g., u and v) to represent the absence of a link and
uppercase Latin letters (e.g., U and V) to denote its presence. Each
pair of nodes can be in one of four states: (i) no edge between the
pair of nodes in layer $\mathcal{U}$ or layer $\mathcal{V}$, denoted
\sstate{\rm u}{\rm v}, (ii) presence of edge in layer $\mathcal{U}$
only, denoted \sstate{\rm U}{\rm v}, (iii) presence of an edge only in
layer $\mathcal{V}$, denoted \sstate{\rm u}{\rm V}, (iv) presence of
an edge in both layers $\mathcal{U}$ and $\mathcal{V}$, denoted
\sstate{\rm U}{\rm V}.
\begin{figure}
\begin{center}
\includegraphics[width=0.48\linewidth,clip]{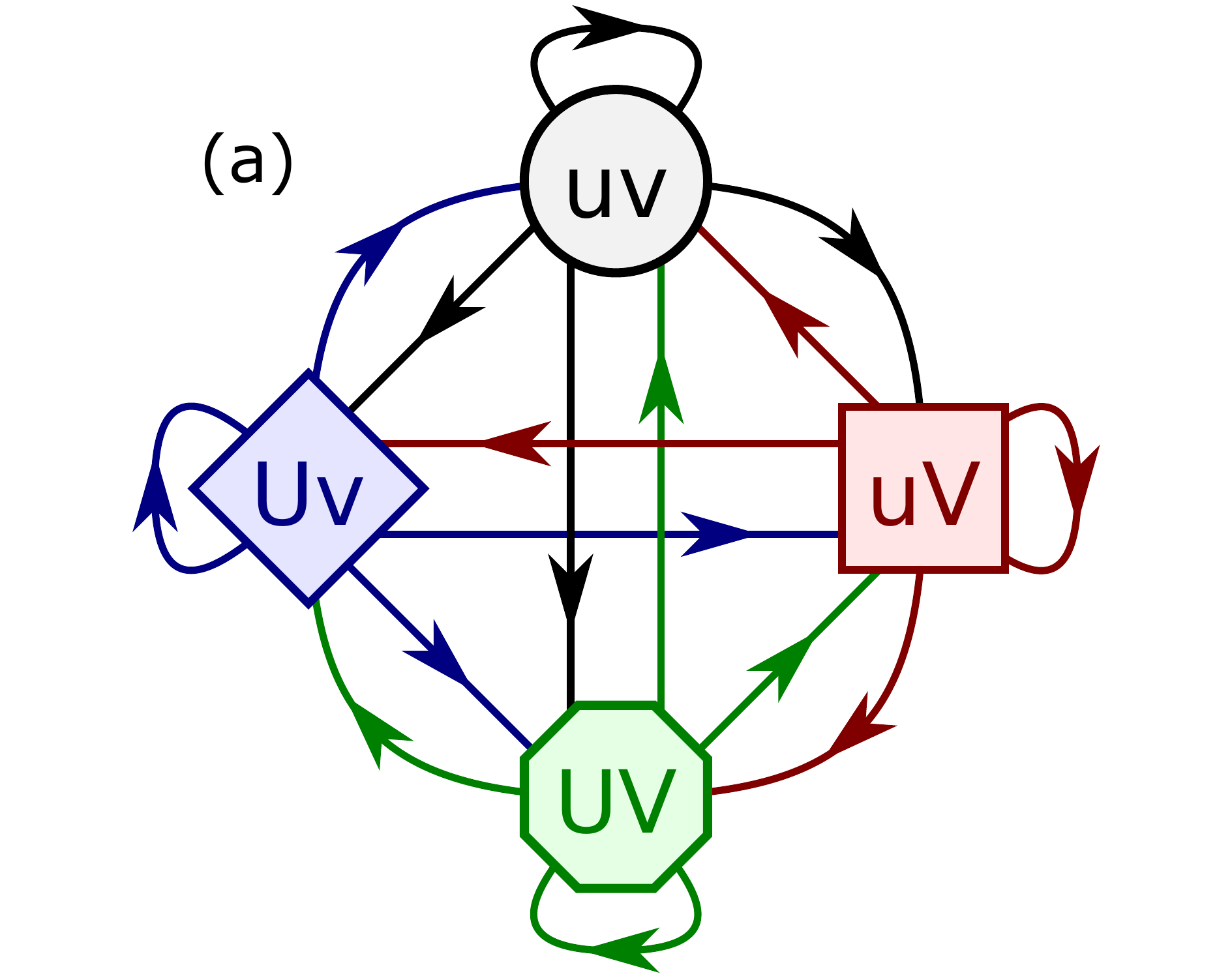}
\includegraphics[width=0.48\linewidth,clip]{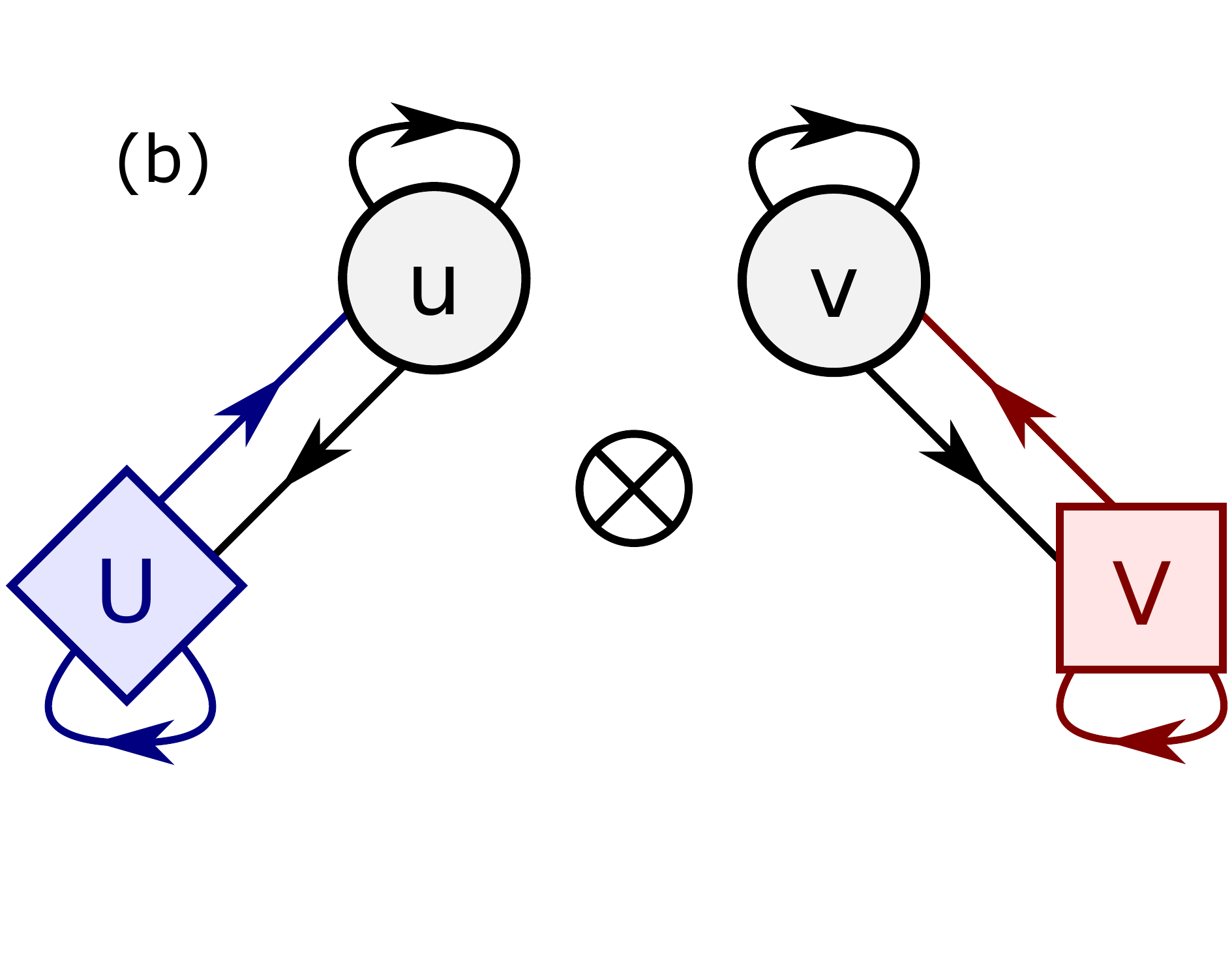}
\caption{(a) Markov chain representation of a multiplex networks with
  two layers $\mathcal{U}$ and $\mathcal{V}$. A lower case letter
  represents absence of a link in the layer and the upper case letter
  represents a presence of a link in the layer. Probabilities that are
  on edges of the same color sum to one. (b) Markov chain
  representation for the edge dynamics on two layers $\mathcal{U}$ and
  $\mathcal{V}$ of a multiplex, assuming they are independent of each
  other. The product of these two Markov chains serves as a null model
  for estimating the presence of spillover.\label{mc1}}
\end{center}
\end{figure}

We construct a Markov chain describing the dynamics of the edges
between a typical pair of nodes in the multiplex network by
considering two consecutive snapshots in time, $t$ and
$t+1$. Formally, this is what we call the \namemc. Since a pair of
nodes can be in one of four possible states in each time step, there
are $16$ possible transitions in this \namemc. Figure~\ref{mc1} (a)
shows such a multiplex Markov chain for a network with two layers,
$\mathcal{U}$ and $\mathcal{V}$. There is a probability associated
with each transition that gives the chance of moving to a state at
time $t+1$ given a state at time $t$, which we refer to as the
transition parameter. These transition parameters are estimated from
the data. Using a Bayesian perspective (see methods) we estimate the
underlying probability distribution for these parameters.

As explained above, there are $16$ possible transitions (including
staying in the same state). Consider every pair of nodes at time $t$
and time $t+1$ and put them in one of $16$ possible bins, each
corresponding to a transition from a state at time $t$ to a state at
time $t+1$. Denote the number of pairs of nodes in each of these bins
using $\mathcal{N}_{ \sstate{\mu}{\nu} \to
  \sstate{\mu^\prime}{\nu^\prime} }$ where \sstate{\mu}{\nu} and
\sstate{\mu^\prime}{\nu^\prime} take values from the set $\mathcal{S}
\equiv$ \{\sstate{\rm u}{\rm v}, \sstate{\rm U}{\rm v}, \sstate{\rm
  u}{\rm V}, \sstate{\rm U}{\rm V}\}. Assuming a uniform prior for the
transition parameters, the mean value of each transition parameter,
$p_{\sstate{\mu}{\nu} \to \sstate{\mu^\prime}{\nu^\prime}}$, can be
obtained by normalizing the counts with the total number of
transitions leaving from a particular state (see methods). The
following expression provides the mean value for the transition
parameter based on the data observed
\begin{equation}
p^{ }_{\sstate{\mu}{\nu} \to \sstate{\mu^\prime}{\nu^\prime}} = \frac{1 + \mathcal{N}_{\sstate{\mu}{\nu} \to \sstate{\mu^\prime}{\nu^\prime}}}{ 4 + \displaystyle \sum_{\sstate{\mu^\prime}{\nu^\prime} \in \mathcal{S}} \mathcal{N}_{\sstate{\mu}{\nu} \to \sstate{\mu^\prime}{\nu^\prime}}}.
\label{eq:ptl}
\end{equation}
In the absence of any data, the above expression results in each
transition parameter having a value of $1/4$, thus reflecting our
assumption of a uniform prior. One measure of the degree of
uncertainty associated with the value obtained above is the standard
deviation, $\sigma^{ }_{ \sstate{\mu}{\nu} \to
  \sstate{\mu^\prime}{\nu^\prime} }$, of the underlying distribution
of the transition parameter. The details of how to obtain these values
is described in the methods section.

To distinguish true correlations from sheer chance, we construct a
null model to compare with the transition parameters of the
{\namemc}. The null model is defined as a Cartesian product of two
Markov chains, one for each layer. This definition ensures that, when
the two layers are truly independent, the transition parameters from
the null model will be exactly the same as those from the {\namemc}.
We first describe how to construct a Markov chain for a single
layer. For a given layer $\mathcal{U}$ of a multiplex network, a pair
of nodes can exist in one of two possible states: (i) no link exists
between the pair, denoted \slstate{\rm u}, and (ii) a link exists
between the pair, denoted \slstate{\rm U}. Similar to the Multiplex
Markov chain, we can obtain the probability of transition for the
single layer from the counts of the transitions
($\mathcal{N}_{\slstate{\mu} \to \slstate{\mu^\prime}}$; $
\slstate{\mu}, \slstate{\mu^\prime} \in \{ \slstate{\rm u},
\slstate{\rm U} \}$) from two consecutive time steps as follows
\begin{equation}
p^{ }_{\slstate{\mu} \to \slstate{\mu^\prime}} =  \frac{1 + \mathcal{N}_{\slstate{\mu} \to \slstate{\mu^\prime}}}{2 + \displaystyle \sum_{\slstate{\mu^\prime} \in \{\slstate{\rm u}, \slstate{\rm U} \}} \mathcal{N}_{\slstate{\mu} \to \slstate{\mu^\prime}}},
\label{eq:psl}
\end{equation}
where $\mathcal{N}_{\slstate{\mu} \to \slstate{\mu^\prime}}$ is the
number of pairs that go from the state \slstate{\mu} at time $t$ to
the state \slstate{\mu^\prime} at time $t+1$. The transition
  probability of the null model is a product of the transition
  probabilities for the respective transitions in the single layer
  Markov chains and can be expressed as
\begin{equation}
p^{\rm null}_{\sstate{\mu}{\nu} \to \sstate{\mu^\prime}{\nu^\prime}} = p^{ }_{\slstate{\mu} \to \slstate{\mu^\prime}} p^{ }_{\slstate{\nu} \to \slstate{\nu^\prime}}.
\end{equation}
The corresponding spread in the distribution of the transition
parameters of the null model can be obtained using the standard way in
which parameters involving a product of Gaussian distributions are
determined (see the methods section for details).

Having obtained the distribution for the transition parameters of the
{\namemc} and the null model, we can examine the extent to which they
overlap. One way to assess such overlaps is to obtain the confidence
intervals associated with the estimated parameters. We say that there
is positive (negative) dynamical spillover when the confidence
interval of the transition parameter of the Multiplex Markov chain
lies entirely above (below) the confidence interval of corresponding
transition parameter of the null model (see methods section for
details).

The code to construct the Multiplex Markov chain and the null model
described in this section can be found on GitHub~\cite{mmc-code}.

\subsection{Networks of nations}
\label{network-of-nations}
\begin{figure}
\begin{center}
\includegraphics[width=0.48\linewidth,clip]{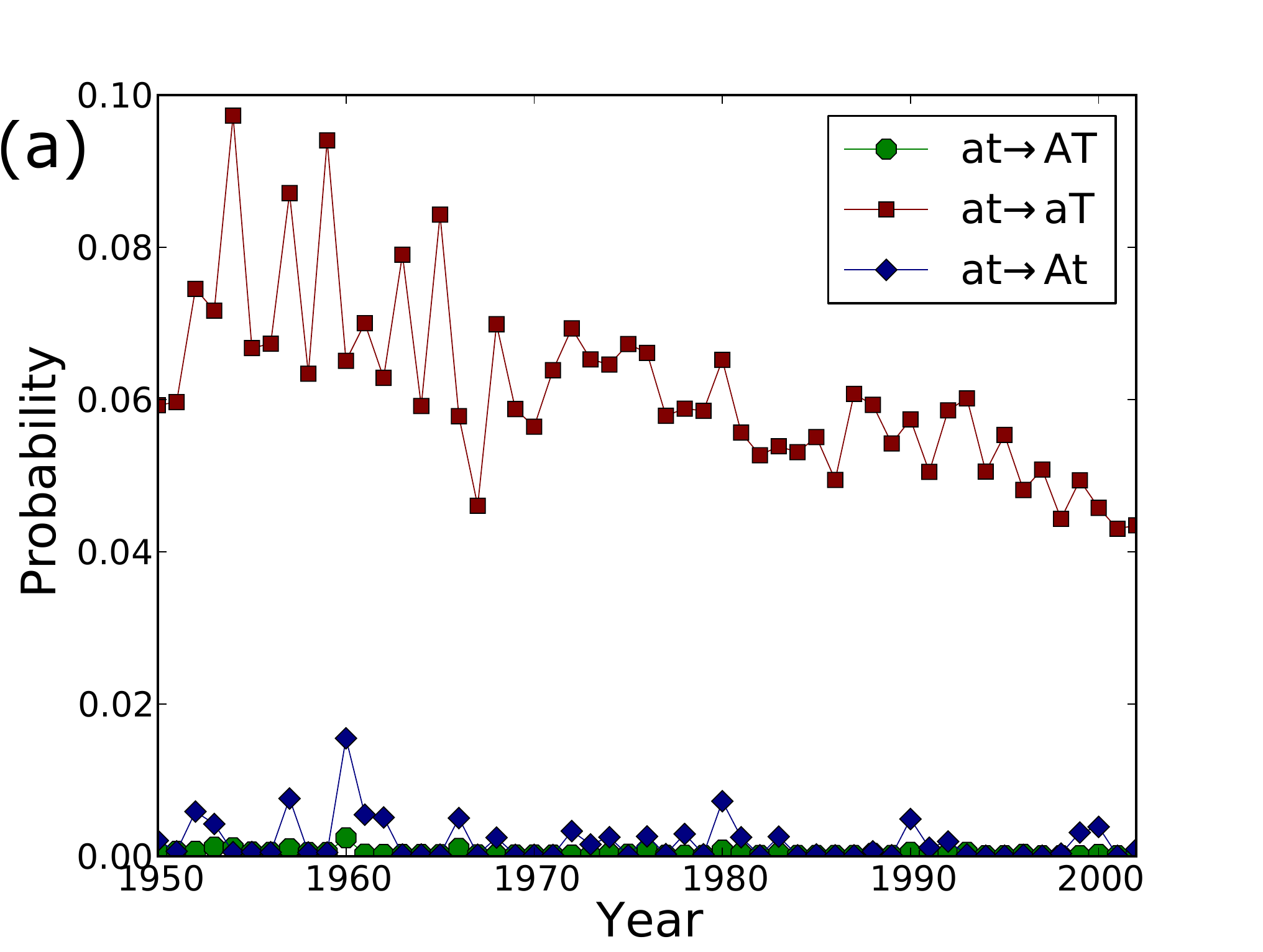}
\includegraphics[width=0.48\linewidth,clip]{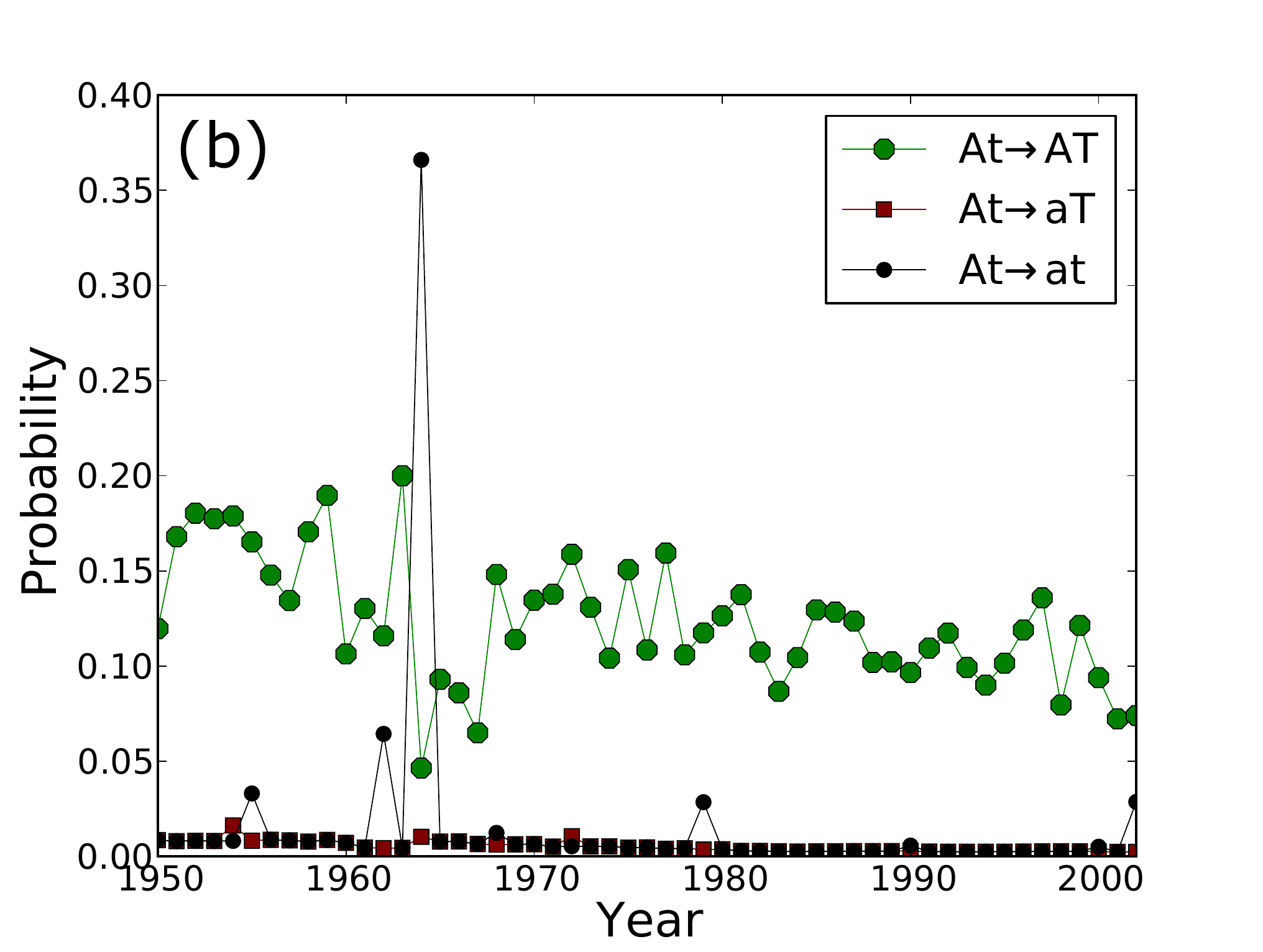}
\includegraphics[width=0.48\linewidth,clip]{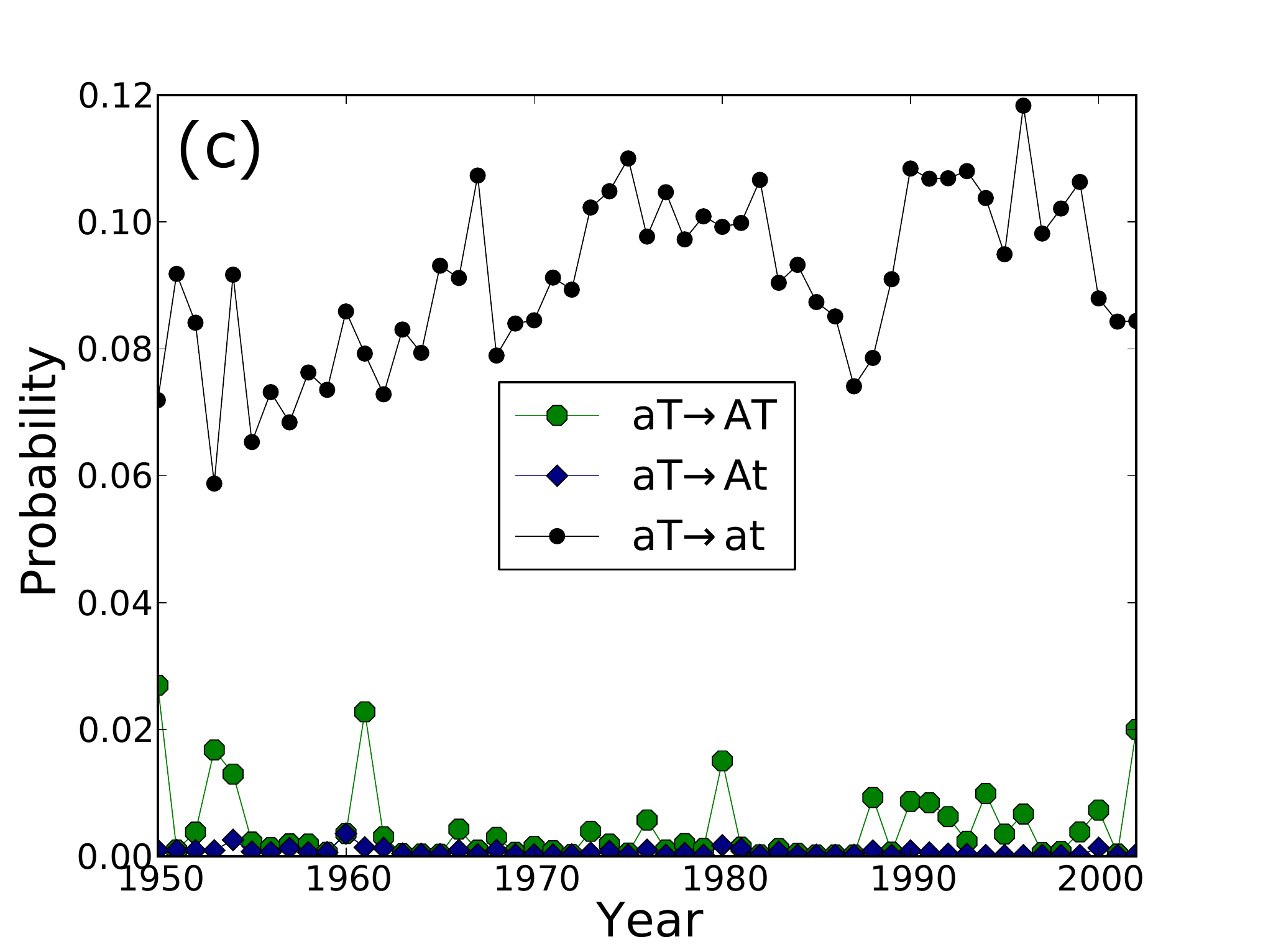}
\includegraphics[width=0.48\linewidth,clip]{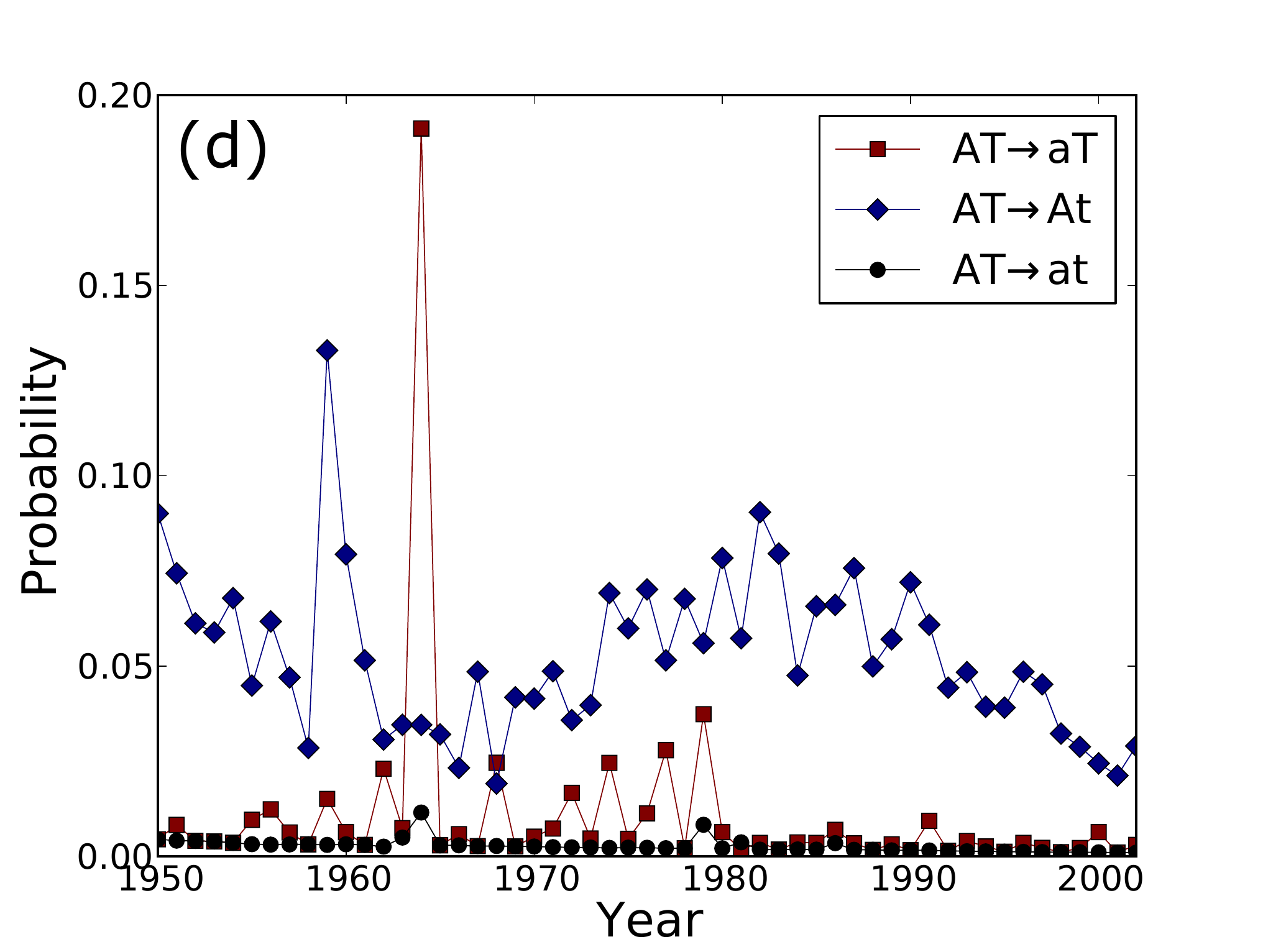}
\caption{Probability of transition obtained from the Multiplex Markov
  chain for each year starting from (a) state with no alliance or
  trade (\sstate{\rm at}), (b) state with only alliance (\sstate{\rm
    At}), (c) state with only trade (\sstate{\rm aT}) and (d) state
  with both trade and alliance (\sstate{\rm AT}). The shape of the
  marker indicates the state to which the transition leads to. Note
  that the scale of the y-axis is different between the
  panels.\label{temporal-prob}}
\end{center}
\end{figure}

Different relationships between countries of the world, such as trade,
alliances, conflicts, etc., can be represented as a multiplex network
with each relationship in a particular layer. A question that is of
particular interest to political scientists who study these systems is
whether international trade and alliances between nations influence
each other. Should there exist a relationship between the two, they
seek to understand if such influence is directional in nature. In this
section we demonstrate how the {\namemc} model may be used to analyze
such a multiplex network and provide key insights to help answer these
questions. Our networks of nations data consist of yearly snapshots of
trade and alliances between nations of the world, starting from the
year $1950$ up to the year $2003$. The edges in this multiplex network
are alliances (denoted $\mathcal{A}$) and trade (denoted
$\mathcal{T}$) which form two layers of this multiplex network. Using
the data from each year, we construct the Multiplex Markov chain which
results in $16$ transition parameters (since there are four states and
there are four transitions originating from each of them) for each
pair of consecutive years. Figure~\ref{temporal-prob} (a--d) shows the
time evolution of these transition parameters. We can see that the
transition parameters are similar for most years except a few. We
perform a principal component analysis (PCA) to examine how similar
the transition parameters for the different years are.

\begin{figure}
\begin{center}
\includegraphics[width=0.48\linewidth,clip]{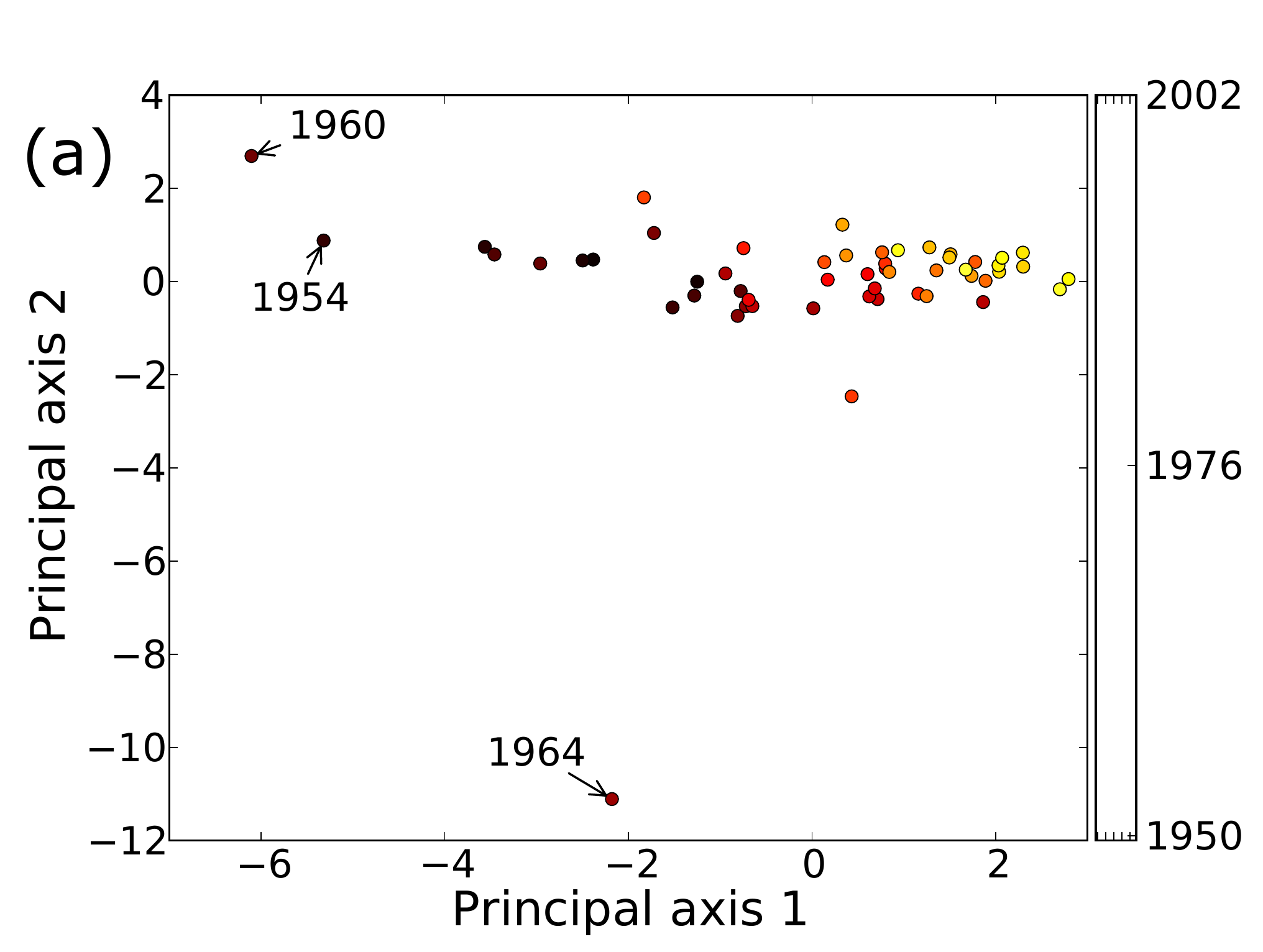}
\includegraphics[width=0.48\linewidth,clip]{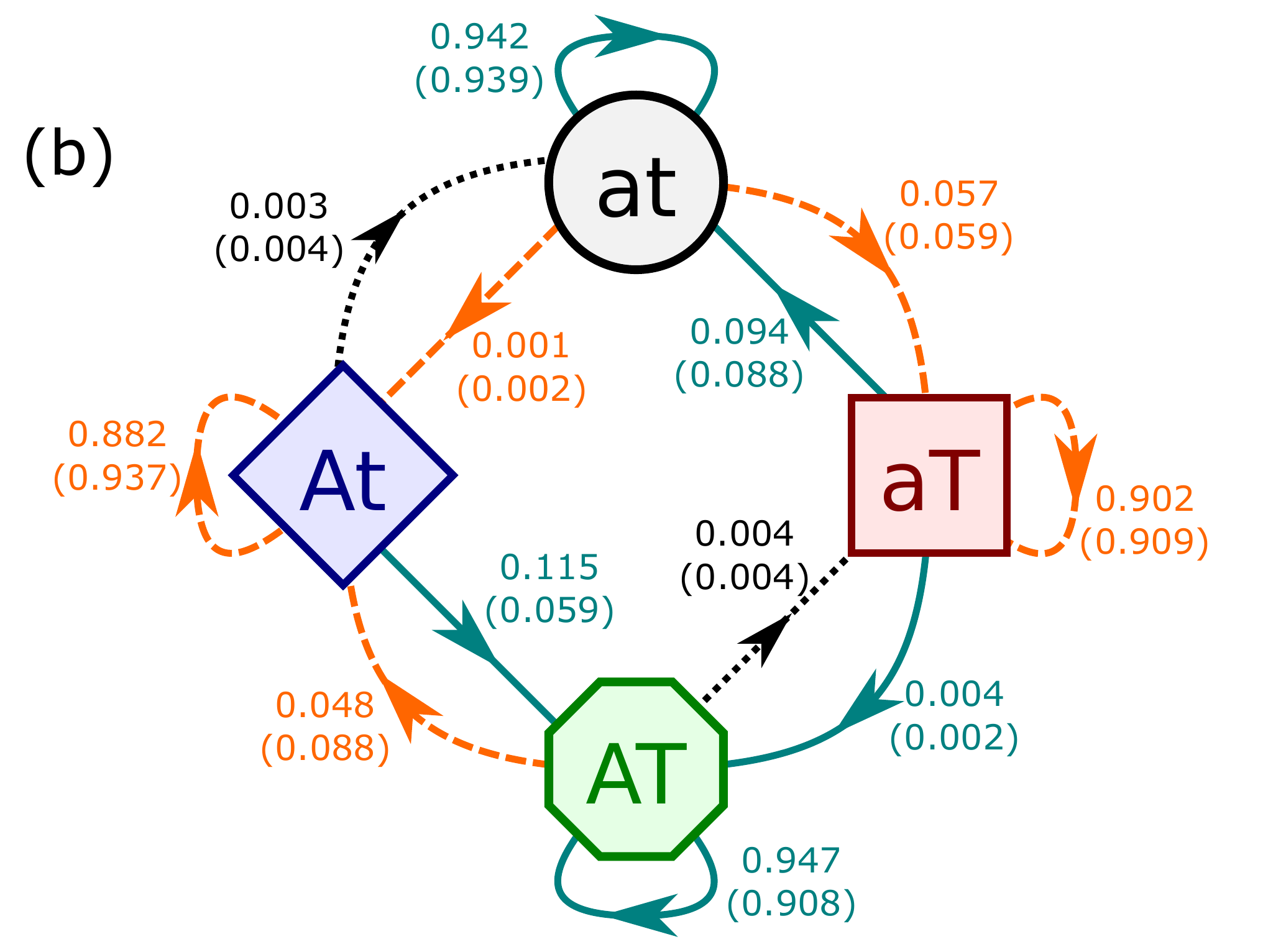}
\caption{(a) PCA of the yearly transition parameters. Note that with
  the exception of a few outliers most points appear to be
  similar. The color of the points correspond to the time period (b)
  Multiplex Markov chain obtained by aggregating the transition
  probabilities for several years. The value of the probability is
  shown on the arrows with the probability obtained from the null
  model within parenthesis. Transition parameters having value less
  than $1 \times 10^{-3}$ are not shown. Solid (dashed) lines
  represent transitions with positive (negative) dynamical
  spillover. Dotted lines show transitions without dynamical
  spillover.\label{mc-trade-alliance}}
\end{center}
\end{figure}

The principal component analysis is performed with only $12$ of the
$16$ transition parameters as the features, since the sum of all
transition parameters leaving a state is equal to one (and hence not
independent). Principal component analysis gives us an ordered list of
a linear combination of the $12$ features. This list is in the
descending order of the amount of variation the linear combinations
accounts for. Figure~\ref{mc-trade-alliance} (a) shows the two
principal component axes that explain the most of the variation found
in the data. Every point on the figure represents the transition
parameters obtained for a given year. The figure shows a trend where
the data points at a later time period have a larger value for
principal axis 1. The reason is as follows. The major contribution to
this principal component axis comes from the transitions \sstate{\rm
  A}{\rm t} $\to$ \sstate{\rm a}{\rm t}, \sstate{\rm A}{\rm T} $\to$
\sstate{\rm a}{\rm t} and \sstate{\rm A}{\rm T} $\to$ \sstate{\rm
  a}{\rm T}, all of which have a negative contribution to the first
principal component axis. In addition, these transitions, on an
average, have a lower value for later time periods when compared to
earlier time periods, suggesting that breaking alliance has become
less likely over time. Taken together, these two facts help explain
the trend observed in the PCA. It is important to note that actual
change in the value of these transition parameters between different
time periods is smaller than $0.01$, but this change is captured and
amplified by the PCA because these transition parameters have a large
variation in their values. This variation can be seen in
Fig.~\ref{temporal-prob} (b), (d) where these three transitions are
close to zero for most years but have a ``spiky'' behavior for a few
years.

Further, we can see that except for the years $1954$, $1960$
and $1964$, most of the years are quite similar and may be aggregated
together. Further, we have verified that including these three outlier
years slightly changes the numerical values of the parameters but does
not affect any aspect of the results discussed below.

The Markov chain obtained by aggregating the counts for the
transitions over all the years except $1954$, $1960$ and $1964$ is
shown in Fig.~\ref{mc-trade-alliance} (b). Note that transition
parameters with values less than $1 \times 10^{-3}$ are excluded from
the figure. We first list observations we can make based only on the
transition parameters of the {\namemc}. In all cases, the pairs of
nodes in the network have a very high probability of staying in the
same state they were present in during the previous year. Consider two
nations that do not have a link in either the alliance or trade layer
at this moment (\sstate{\rm a}{\rm t}) but end up in a state with a
link in both layers (\sstate{\rm A}{\rm T}) at some point in the
future. The Multiplex Markov chain obtained from the data tells us
that the states are most likely to first get a link in the trade
layer, and then form a link in the alliance layer (in addition to
having trade). Also, if two states have an alliance to begin with,
they have a very high likelihood of quickly gaining a trading edge in
addition to having an alliance.

Having examined the transition parameters obtained from the data, we
now focus our attention on dynamical spillover between the two layers
by looking for differences between the transition parameters from the
{\namemc} and the null model. We find that, at this aggregate level,
the null model does a good job in capturing the probabilities for many
transitions. The most notable exception is the transition parameter
from \sstate{\rm A}{\rm t} to \sstate{\rm A}{\rm T} which is much
higher when compared to that obtained from the null model. This
suggests that the probability that two states establish a trade
relationship (with each other) once they have formed an alliance is
considerably higher when compared to the two events happening
independent of each other. Another difference between the multiplex
Markov chain and the null model is the probability of leaving state
\sstate{\rm A}{\rm T} and going back to a state of only alliance
(\sstate{\rm A}{\rm t}). This probability is approximately twice as
small in the observed data when compared to the null model.

These results shed light on a longstanding debate in the political
science literature on the precise nature of the relationship between
alliances and international trade. Employing multiple regression
models, \citet{manbro97} examined the effects of alliances on
trade. They found that the interaction between alliance and trade
agreements at year $t$ increases trade flow between the same countries
at year $t+1$. Additional studies by \citet{gow95},
\citet{gowman93,gowman04}, \citet{lon03}, and \citet{lonlee06}, among
others, are consistent with this basic finding. Nonetheless,
\citet{morsiv98}, using a regression model with a different set of
independent variables and alternative model specifications, suggest
that alliances do not have a robust and consistent effect on
trade. From Fig.~\ref{mc-trade-alliance} (b), we see that the
probability of forming a link in the trade layer when there is no
alliance ($p_{\sstate{\rm a}{\rm t} \to \sstate{\rm a}{\rm T}}$) is
about twice as small when compared to the probability of forming a
trade link when an alliance is already present ($p_{\sstate{\rm A}{\rm
    t} \to \sstate{\rm A}{\rm T}}$). We can also see that breaking of
a trade link is much more likely when an alliance is absent when
compared to when an alliance is present ({\ie} $p^{ }_{\sstate{\rm
    a}{\rm T} \to \sstate{\rm a}{\rm t}} > p^{ }_{\sstate{\rm A}{\rm
    T} \to \sstate{\rm A}{\rm t}}$). These observations are consistent
with the adage in political science that ``trade follows the flag'',
\ie, the forging of economic relations---both during the Cold War era
(1950-1990) and the post-Cold War era (1991-2003)---seems to have been
affected by the presence of security cooperation between states.

There have also been other studies in political science, such as
\citet{for10} and Maoz~\cite{maoz2010}, that study the impact trade
has on alliances between countries. These studies show that trade
increases the likelihood of alliance formation and reduces the
probability of alliance dissolution. From Fig.~\ref{mc-trade-alliance}
(b) we see that the probability of forming an alliance when trade is
present ($p_{\sstate{\rm a}{\rm T} \to \sstate{\rm A}{\rm T}}$) is
about four times as large as the probability of forming an alliance
when no trade is present ($p_{\sstate{\rm a}{\rm t} \to \sstate{\rm
    A}{\rm t}}$). This indicates that having trade between two
countries helps in the formation of an alliance between them. However,
we are unable say anything conclusive about the effect of trade on the
dissolution of alliances. Based on the above results we can see that
our method using the Multiplex Markov chain paints a much richer
picture and explores potential causal pathways that are found in the
data.

\subsection{Network of open source software developers}
\begin{figure}
\begin{center}
\includegraphics[width=0.48\linewidth,clip]{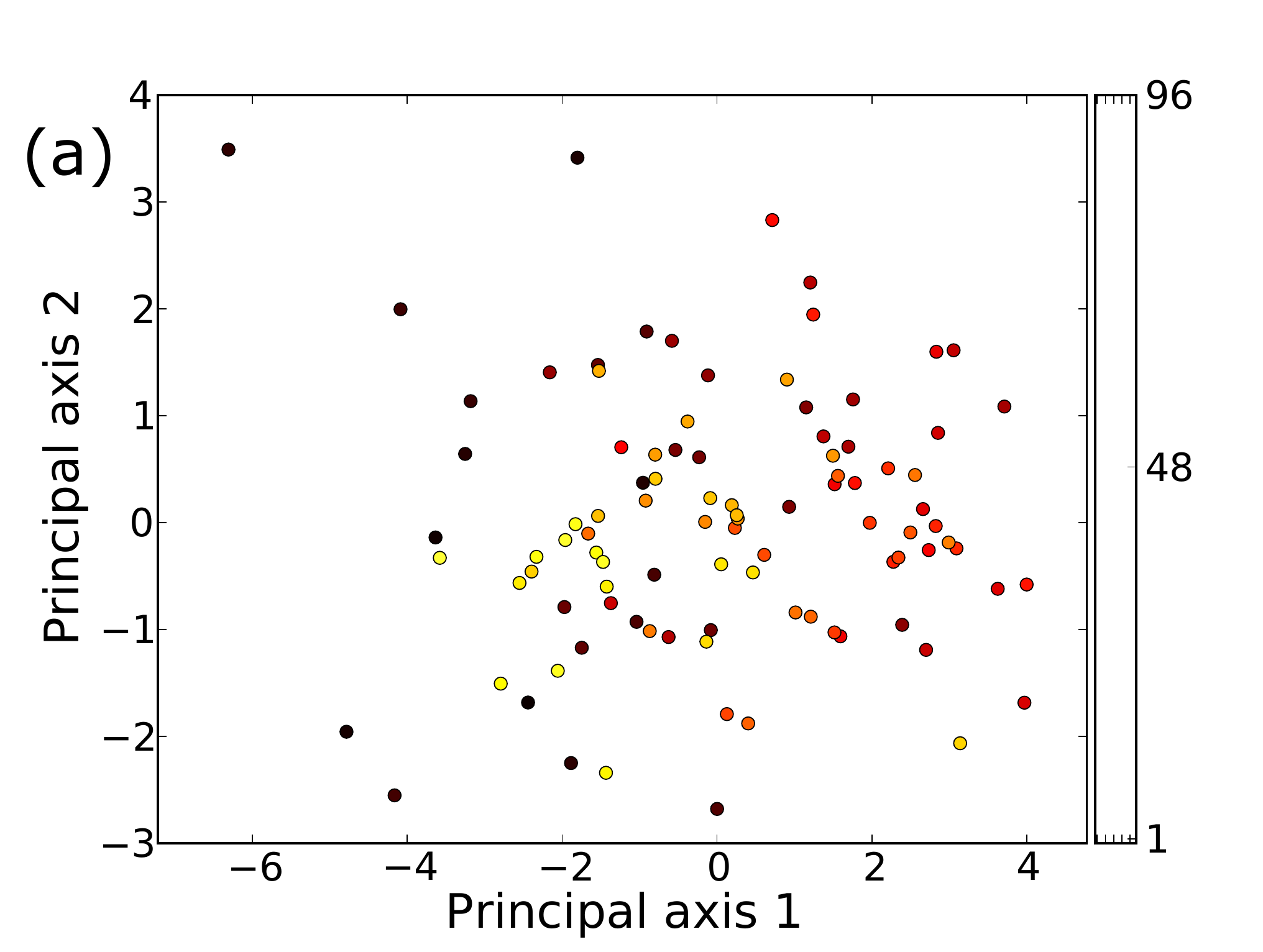}
\includegraphics[width=0.48\linewidth,clip]{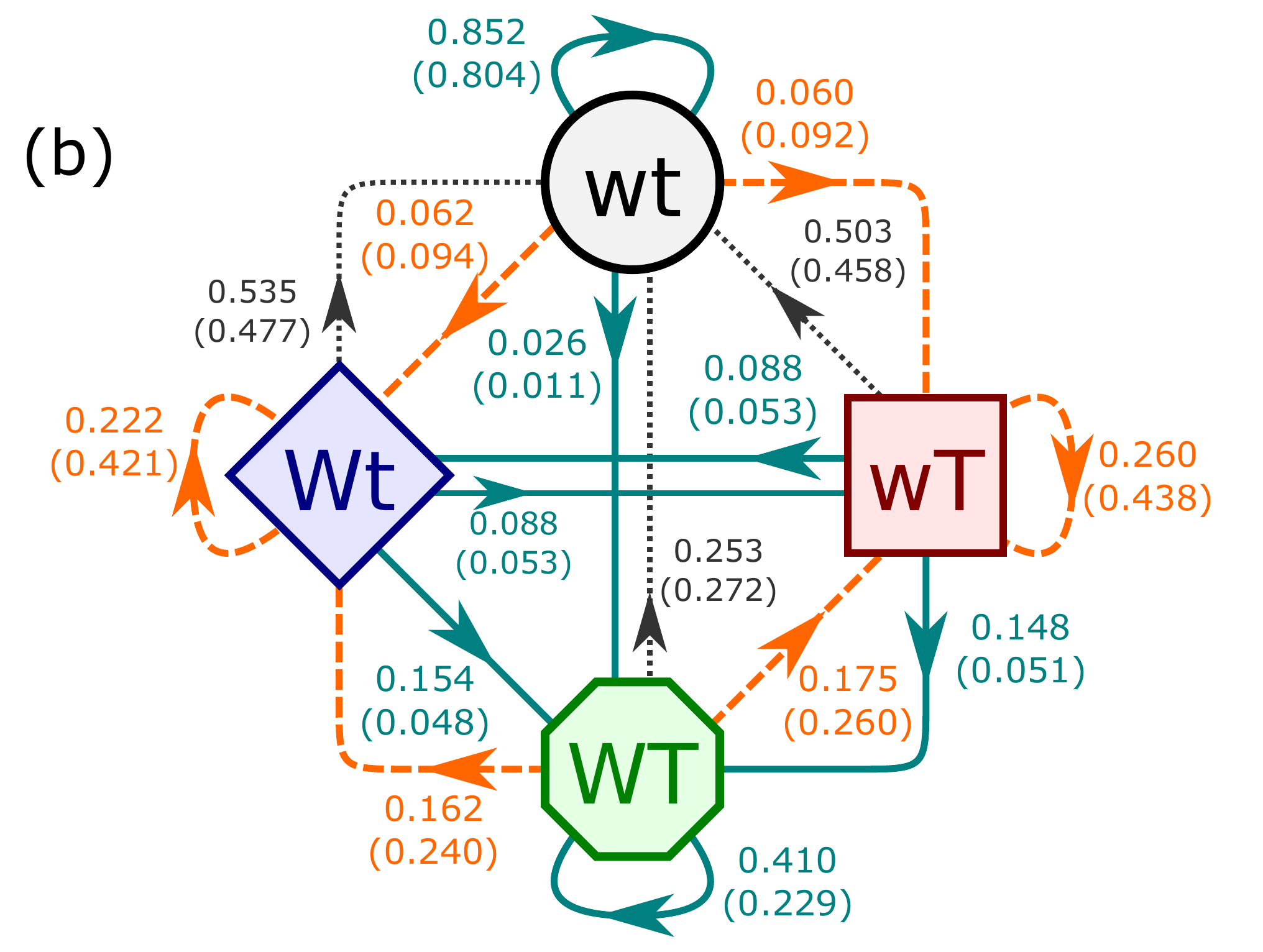}
\caption{(a) PCA of the transition parameters of subsequent 4-week
  intervals. The color of the points correspond to the time period (b)
  Multiplex Markov chain obtained by aggregating the transition
  parameters for all 4-week time intervals. The value of the
  transition parameter is shown on the arrows along with the
  transition parameter obtained from the null model within
  parenthesis. Solid (dashed) lines represent transitions with
  positive (negative) dynamical spillover. Dotted lines show
  transitions without dynamical spillover.\label{acc-mc}}
\end{center}
\end{figure}

Open Source Software (OSS) projects are great examples of how people
come together without a formal organization to collaborate and perform
complex tasks. The existence of records of the collaboration
activities among developers of OSS projects makes them ideal test-beds
for helping uncover the more general question about how cooperation
emerges in societies. A problem that has received considerable
attention of researchers in this area is the emergence of
collaboration among developers and how it is influenced by social ties
between them~\cite{gharehyazie2014, nakakoji2002, ducheneaut2005,
  xuan2012, xuan2014}. Here, we apply the tool developed in
Sec.~\ref{mc-method} to a particular OSS project to show how our model
may help untangle the relationship between communication and
collaboration networks.

We study ``Apache Axis2/Java'', an open source software project that
is part of the Apache software foundation~\cite{apache}. Our data
consists of email communications to a mailing list and co-commits by
developers starting from $2^{\rm nd}$ September 2004 to $19^{\rm th}$
March 2013. The project had $68$ developers who contributed code at
some point during our period of observation. The nodes in the network
are developers and the two layers are email (T, for talk) and
collaboration (W, for work). We use co-commit, \ie, two developers
editing the same file, to indicate collaboration between them. We
consider a link to be present between two nodes in the respective
layer if that activity occurred at least once in a particular four
week interval. Therefore our snapshots in time correspond to
non-overlapping four week time periods. Different four week time
periods have different number of developers. This number ranges
between $2$ and $29$ developers. Due to these number being so small,
the data is quite noisy and there is not much we can infer from the
raw time series. Therefore, we aggregate the counts for the
transitions of the {\namemc} to discern if there exists dynamical
spillover between the email and co-commit networks. Before
aggregating, similar to Sec.~\ref{network-of-nations}, we perform a
PCA with $12$ of the $16$ transition parameters as the
features. Figure~\ref{acc-mc} (a) shows the results of the PCA, where
the axes shown are the top two in terms of the amount of variation in
the data the axes account for. There seems to be a single cluster of
points and no obvious trend in the PCA. This justifies aggregating the
counts for the transitions from the different 4-week intervals.

Figure~\ref{acc-mc} (b) shows the results aggregated over all the time
periods. The project has a much higher value for the probability of
remaining in the state of having both email and co-commit
($\sstate{\rm W}{\rm T} \to \sstate{\rm W}{\rm T}$) when compared to
the null, indicating that the correlations between the two layers may
play an important role in sustaining collaborations between
developers. Moreover, transitions from both $\sstate{\rm W}{\rm t}$
and $\sstate{\rm w}{\rm T}$ states to the state $\sstate{\rm W}{\rm
  T}$ exhibit positive dynamical spillover showing that both
communication and collaboration between developers can lead to a
sustaining state where developers both collaborate and communicate
with each other. Studies by Xuan et al~\cite{xuan2012, xuan2014}, on
open source software using different techniques, agree qualitatively
with the results we obtain.

\section{Discussion}
In this article we construct a model based on Markov chains to
ascertain the existence of dynamical spillover in real world multiplex
networks. By constructing an appropriate null model, we establish how
to distinguish the correlations that arise from sheer randomness and
those that arise from non-trivial dynamics present in the system; we
refer to the non-random correlations as dynamical spillover. In
addition, we apply the tool to two real world datasets: (i) trade and
alliance between nation-states, and (ii) email and co-commit network
of an open source software project. We show that these networks
exhibit non-trivial co-evolution among the layers of their respective
multiplex networks, and tease out some potential causal pathways. The
approach described in this article studies the presence of
correlations at the edge level (aggregated over all pairs). A question
that future works may focus on is how these correlations affect
network structures such as the presence of a core-periphery structure
or communities. Future studies could also extend this technique to
include neighborhood of the edges and systematically study the effect
of such correlations on the dynamics on the network.

\section{Methods}

\subsection{Uncertainty in transition parameters of a Markov chain}

Both the {\namemc} and the null model require that we estimate the
transition parameters of a Markov chain. First consider a transition
parameter $p^{}_{\slstate{\mu} \to \slstate{\mu^\prime}}$ for a single
layer. Using the prior that the parameter is uniformly taken from the
interval $[0,1]$, a full Bayesian inference would update on the
information $\mathcal{N}^{}_{\slstate{\mu} \to \slstate{u}}$ and
$\mathcal{N}^{}_{\slstate{\mu} \to \slstate{U}}$ to yield a beta
distribution~\cite{sivia2006data}. The mean of this distribution is
the value given in Eq.~\eqref{eq:psl}. The standard deviation of this
beta distribution is a measure of variance in the estimated transition
parameter. The standard deviation is obtained using the formula
\begin{equation}
\sigma^{ }_{\slstate{\mu} \to \slstate{\mu^\prime}} =  \sqrt{\frac{p^{ }_{\slstate{\mu} \to \slstate{\mu^\prime}}(1 - p^{ }_{\slstate{\mu} \to \slstate{\mu^\prime}})}{\displaystyle 3 + \sum_{\slstate{\mu^\prime} \in \{\slstate{\rm u}, \slstate{\rm U} \}} \mathcal{N}_{\slstate{\mu} \to \slstate{\mu^\prime}}}}.
\end{equation}

When two layers are involved in the transition, a full Bayesian
inference with a uniform prior would yield a Dirichlet (i.e.,
multivariate beta) distribution of order $4$ with the mean value of
the different variates given by Eq.~\eqref{eq:ptl}. The standard
deviation associated with the variates of such a Dirichlet
distribution of order $4$ is given by
\begin{equation}
\sigma^{ }_{ \sstate{\mu}{\nu} \to \sstate{\mu^\prime}{\nu^\prime} } = \sqrt{\frac{p^{ }_{\sstate{\mu}{\nu} \to \sstate{\mu^\prime}{\nu^\prime}}(1 - p^{ }_{\sstate{\mu}{\nu} \to \sstate{\mu^\prime}{\nu^\prime}})}{ 5 + \displaystyle \sum_{\sstate{\mu^\prime}{\nu^\prime} \in \mathcal{S}} \mathcal{N}_{\sstate{\mu}{\nu} \to \sstate{\mu^\prime}{\nu^\prime}}}}.
\end{equation}
Note that our approach does not need the values for the covariances.

\subsection{Uncertainty of the null model}
The spread in the distribution of the transition parameter of the null
model will depend on the spread in the distribution of transition
parameters of the single layer Markov chains from which we construct
our null model. Once we have determined the standard deviation of the
transition parameters of the single layer Markov chains, we can use
them to estimate the spread in the transition parameters of the null
model. Since the transition parameters are sufficiently far away from
$0$ and $1$, and since the counts from which we estimate the
transition parameters are sufficiently high (of the order of $10^3$),
the beta distribution can be approximated as a Gaussian
distribution. This leads to the following expression for standard
deviation of the transition parameters of the null model.
\begin{equation}
\sigma^{\rm null}_{\sstate{\mu}{\nu} \to \sstate{\mu^\prime}{\nu^\prime} } \approx p^{\rm null}_{\sstate{\mu}{\nu} \to \sstate{\mu^\prime}{\nu^\prime}} \sqrt{ \left(\frac{\sigma^{ }_{\slstate{\mu} \to \slstate{\mu^\prime}}}{p^{ }_{\slstate{\mu} \to \slstate{\mu^\prime}}} \right)^2 + \left(\frac{\sigma^{ }_{\slstate{\nu} \to \slstate{\nu^\prime}}}{p^{ }_{\slstate{\nu} \to \slstate{\nu^\prime}}} \right)^2}.
\end{equation}
In obtaining the above approximation we have assumed that the term
involving the product $\sigma^{ }_{\slstate{\mu} \to
  \slstate{\mu^\prime}} \sigma^{ }_{\slstate{\nu} \to
  \slstate{\nu^\prime}}$ is small when compared to other terms
present. Our assumption is justified since $\sigma^{ }_{\slstate{\mu}
  \to \slstate{\mu^\prime}} \ll 1$ and $\sigma^{ }_{\slstate{\nu} \to
  \slstate{\nu^\prime}} \ll 1$ so their product is even smaller. In
cases where the data is not sufficient to justify the assumptions made
above we can compute the standard deviation by numerically sampling
the two beta distributions associated with the null model.

\subsection{Dynamical spillover}
We say that dynamical spillover exists between layers of a multiplex
network if at least one of the transition parameters of the {\namemc}
is ``substantially'' different from that of the null model. We can say
that the difference is substantial when the confidence intervals
associated with the transition parameter of the {\namemc} and
corresponding transition parameter of the null model do not
overlap. Further, when the confidence interval of a transition
parameter for the {\namemc} lies entirely above (below) the confidence
interval of the corresponding transition parameter of the null model
we refer to it as positive (negative) dynamical spillover. In this
work, we choose the $99 \%$ confidence interval for both the {\namemc}
and the null model.  Note that the transition probabilities from the
{\namemc} and the null model are positively correlated and hence this
is a conservative estimate of whether the transition parameter
obtained from the {\namemc} and null model are indeed considerably
different.

The Python code to discern the existence of dynamical spillover by
constructing the {\namemc} and the null model can be found on
Github~\cite{mmc-code}.

\subsection{Data}
The data on international alliances is derived from the ATOP
project~\cite{leeds2005}. Trade networks are derived from a
combination of two datasets on international trade: the Correlates of
War (COW) bilateral trade dataset~\cite{barbieri2009}, and the
Gledistch trade dataset~\cite{gleditsch2002}.  These datasets are
combined to reduce the amount of missing data, but they are similar in
all other respects. The trade data is binarized such that an edge
exists if the value of trade between two countries is greater than
$10^{-3}$ of the exporter's GDP. Note the above threshold is not
symmetric between the two countries involved in trade, and hence gives
us a directed network of trade between nations. Neglecting the edge
directionality (i.e., ensuring there is at least one-way trade) yields
the undirected networks studied herein.

The email network of OSS developers is constructed by scrapping online
records of the mailing lists~\cite{oss-email-url}. The co-commit
network is obtained as a projection of the bipartite network of
developers and the files they commit to. Each commit represents an
edge in this bipartite network. The information is publicly available
on the git repository of the OSS project. We consider snapshots in
time aggregated over non-overlapping four week intervals. Unlike the
network of nations, where we know if a country exists in a given year
independent of the trade and alliance relationships, in this data we
are not sure if a developer is actively contributing to the project
and have to infer the existence of the developer based on their
activity either in the email or co-commit network. We consider the
developer to be part of the system if she/he has at least one edge (in
either layer) in the current time step or the next time step. This
tends to underestimate the probability of a developer who is neither
actively committing code nor sending emails to the mailing list will
remain in the same state.

\section{Acknowledgments}
We are grateful to Charlie Brummitt, George Barnett, Kyle Joyce,
Brandon Kinne, Tracy Lin and Qi Xuan for helpful discussions. We thank
Vladimir Filkov, Premkumar Devanbu and Qi Xuan for providing access to
the data on open source software networks. We gratefully acknowledge
support from the US Army Research Laboratory and the US Army Research
Office under MURI award W911NF-13-1-0340, and Cooperative Agreement
W911NF-09-2-0053, the Defense Threat Reduction Agency Basic Research
Grant No. HDTRA1-10-1-0088 and NSF Grant No. ICES-1216048.

%


\begin{thebibliography}{43}%
\makeatletter
\providecommand \@ifxundefined [1]{%
 \@ifx{#1\undefined}
}%
\providecommand \@ifnum [1]{%
 \ifnum #1\expandafter \@firstoftwo
 \else \expandafter \@secondoftwo
 \fi
}%
\providecommand \@ifx [1]{%
 \ifx #1\expandafter \@firstoftwo
 \else \expandafter \@secondoftwo
 \fi
}%
\providecommand \natexlab [1]{#1}%
\providecommand \enquote  [1]{``#1''}%
\providecommand \bibnamefont  [1]{#1}%
\providecommand \bibfnamefont [1]{#1}%
\providecommand \citenamefont [1]{#1}%
\providecommand \href@noop [0]{\@secondoftwo}%
\providecommand \href [0]{\begingroup \@sanitize@url \@href}%
\providecommand \@href[1]{\@@startlink{#1}\@@href}%
\providecommand \@@href[1]{\endgroup#1\@@endlink}%
\providecommand \@sanitize@url [0]{\catcode `\\12\catcode `\$12\catcode
  `\&12\catcode `\#12\catcode `\^12\catcode `\_12\catcode `\%12\relax}%
\providecommand \@@startlink[1]{}%
\providecommand \@@endlink[0]{}%
\providecommand \url  [0]{\begingroup\@sanitize@url \@url }%
\providecommand \@url [1]{\endgroup\@href {#1}{\urlprefix }}%
\providecommand \urlprefix  [0]{URL }%
\providecommand \Eprint [0]{\href }%
\providecommand \doibase [0]{http://dx.doi.org/}%
\providecommand \selectlanguage [0]{\@gobble}%
\providecommand \bibinfo  [0]{\@secondoftwo}%
\providecommand \bibfield  [0]{\@secondoftwo}%
\providecommand \translation [1]{[#1]}%
\providecommand \BibitemOpen [0]{}%
\providecommand \bibitemStop [0]{}%
\providecommand \bibitemNoStop [0]{.\EOS\space}%
\providecommand \EOS [0]{\spacefactor3000\relax}%
\providecommand \BibitemShut  [1]{\csname bibitem#1\endcsname}%
\let\auto@bib@innerbib\@empty
\bibitem [{\citenamefont {Nagurney}\ and\ \citenamefont
  {Dong}(2002)}]{nagurney2002}%
  \BibitemOpen
  \bibfield  {author} {\bibinfo {author} {\bibfnamefont {A.}~\bibnamefont
  {Nagurney}}\ and\ \bibinfo {author} {\bibfnamefont {J.}~\bibnamefont
  {Dong}},\ }\href@noop {} {\emph {\bibinfo {title} {Supernetworks:
  Decision-Making for the Information Age}}}\ (\bibinfo  {publisher} {Elgar,
  Edward Publishing, Incorporated},\ \bibinfo {year} {2002})\BibitemShut
  {NoStop}%
\bibitem [{\citenamefont {D'Agostino}\ and\ \citenamefont
  {Scala}(2014)}]{d2014}%
  \BibitemOpen
  \bibfield  {author} {\bibinfo {author} {\bibfnamefont {G.}~\bibnamefont
  {D'Agostino}}\ and\ \bibinfo {author} {\bibfnamefont {A.}~\bibnamefont
  {Scala}},\ }\href@noop {} {\emph {\bibinfo {title} {Networks of Networks: The
  Last Frontier of Complexity}}}\ (\bibinfo  {publisher} {Springer},\ \bibinfo
  {year} {2014})\BibitemShut {NoStop}%
\bibitem [{\citenamefont {Buldyrev}\ \emph {et~al.}(2010)\citenamefont
  {Buldyrev}, \citenamefont {Parshani}, \citenamefont {Paul}, \citenamefont
  {Stanley},\ and\ \citenamefont {Havlin}}]{buldyrev2010}%
  \BibitemOpen
  \bibfield  {author} {\bibinfo {author} {\bibfnamefont {S.}~\bibnamefont
  {Buldyrev}}, \bibinfo {author} {\bibfnamefont {R.}~\bibnamefont {Parshani}},
  \bibinfo {author} {\bibfnamefont {G.}~\bibnamefont {Paul}}, \bibinfo {author}
  {\bibfnamefont {H.}~\bibnamefont {Stanley}}, \ and\ \bibinfo {author}
  {\bibfnamefont {S.}~\bibnamefont {Havlin}},\ }\href {\doibase
  10.1038/nature08932} {\bibfield  {journal} {\bibinfo  {journal} {Nature}\
  }\textbf {\bibinfo {volume} {464}},\ \bibinfo {pages} {1025} (\bibinfo {year}
  {2010})}\BibitemShut {NoStop}%
\bibitem [{\citenamefont {Leicht}\ and\ \citenamefont
  {D'Souza}(2009)}]{leicht2009}%
  \BibitemOpen
  \bibfield  {author} {\bibinfo {author} {\bibfnamefont {E.}~\bibnamefont
  {Leicht}}\ and\ \bibinfo {author} {\bibfnamefont {R.~M.}\ \bibnamefont
  {D'Souza}},\ }\href@noop {} {\bibfield  {journal} {\bibinfo  {journal} {arXiv
  preprint arXiv:0907.0894}\ } (\bibinfo {year} {2009})}\BibitemShut {NoStop}%
\bibitem [{\citenamefont {Katz}\ and\ \citenamefont {Powell}(1953)}]{katz1953}%
  \BibitemOpen
  \bibfield  {author} {\bibinfo {author} {\bibfnamefont {L.}~\bibnamefont
  {Katz}}\ and\ \bibinfo {author} {\bibfnamefont {J.}~\bibnamefont {Powell}},\
  }\href {\doibase 10.1007/BF02289063} {\bibfield  {journal} {\bibinfo
  {journal} {Psychometrika}\ }\textbf {\bibinfo {volume} {18}},\ \bibinfo
  {pages} {249} (\bibinfo {year} {1953})}\BibitemShut {NoStop}%
\bibitem [{\citenamefont {Boorman}\ and\ \citenamefont
  {White}(1976)}]{boorman1976}%
  \BibitemOpen
  \bibfield  {author} {\bibinfo {author} {\bibfnamefont {S.~A.}\ \bibnamefont
  {Boorman}}\ and\ \bibinfo {author} {\bibfnamefont {H.~C.}\ \bibnamefont
  {White}},\ }\href {http://www.jstor.org/stable/2777009} {\bibfield  {journal}
  {\bibinfo  {journal} {American Journal of Sociology}\ }\textbf {\bibinfo
  {volume} {81}},\ \bibinfo {pages} {pp. 1384} (\bibinfo {year}
  {1976})}\BibitemShut {NoStop}%
\bibitem [{\citenamefont {Hubert}\ and\ \citenamefont
  {Baker}(1978)}]{hubert1978}%
  \BibitemOpen
  \bibfield  {author} {\bibinfo {author} {\bibfnamefont {L.}~\bibnamefont
  {Hubert}}\ and\ \bibinfo {author} {\bibfnamefont {F.}~\bibnamefont {Baker}},\
  }\href {\doibase 10.1007/BF02294087} {\bibfield  {journal} {\bibinfo
  {journal} {Psychometrika}\ }\textbf {\bibinfo {volume} {43}},\ \bibinfo
  {pages} {31} (\bibinfo {year} {1978})}\BibitemShut {NoStop}%
\bibitem [{\citenamefont {Pattison}\ \emph {et~al.}(2000)\citenamefont
  {Pattison}, \citenamefont {Wasserman}, \citenamefont {Robins},\ and\
  \citenamefont {Kanfer}}]{pattison2000}%
  \BibitemOpen
  \bibfield  {author} {\bibinfo {author} {\bibfnamefont {P.}~\bibnamefont
  {Pattison}}, \bibinfo {author} {\bibfnamefont {S.}~\bibnamefont {Wasserman}},
  \bibinfo {author} {\bibfnamefont {G.}~\bibnamefont {Robins}}, \ and\ \bibinfo
  {author} {\bibfnamefont {A.~M.}\ \bibnamefont {Kanfer}},\ }\href {\doibase
  http://dx.doi.org/10.1006/jmps.1999.1261} {\bibfield  {journal} {\bibinfo
  {journal} {Journal of Mathematical Psychology}\ }\textbf {\bibinfo {volume}
  {44}},\ \bibinfo {pages} {536 } (\bibinfo {year} {2000})}\BibitemShut
  {NoStop}%
\bibitem [{\citenamefont {Cai}\ \emph {et~al.}(2005{\natexlab{a}})\citenamefont
  {Cai}, \citenamefont {Shao}, \citenamefont {He}, \citenamefont {Yan},\ and\
  \citenamefont {Han}}]{cai2005}%
  \BibitemOpen
  \bibfield  {author} {\bibinfo {author} {\bibfnamefont {D.}~\bibnamefont
  {Cai}}, \bibinfo {author} {\bibfnamefont {Z.}~\bibnamefont {Shao}}, \bibinfo
  {author} {\bibfnamefont {X.}~\bibnamefont {He}}, \bibinfo {author}
  {\bibfnamefont {X.}~\bibnamefont {Yan}}, \ and\ \bibinfo {author}
  {\bibfnamefont {J.}~\bibnamefont {Han}},\ }in\ \href {\doibase
  10.1007/11564126_44} {\emph {\bibinfo {booktitle} {Knowledge Discovery in
  Databases: PKDD 2005}}},\ \bibinfo {series} {Lecture Notes in Computer
  Science}, Vol.\ \bibinfo {volume} {3721},\ \bibinfo {editor} {edited by\
  \bibinfo {editor} {\bibfnamefont {A.}~\bibnamefont {Jorge}}, \bibinfo
  {editor} {\bibfnamefont {L.}~\bibnamefont {Torgo}}, \bibinfo {editor}
  {\bibfnamefont {P.}~\bibnamefont {Brazdil}}, \bibinfo {editor} {\bibfnamefont
  {R.}~\bibnamefont {Camacho}}, \ and\ \bibinfo {editor} {\bibfnamefont
  {J.}~\bibnamefont {Gama}}}\ (\bibinfo  {publisher} {Springer Berlin
  Heidelberg},\ \bibinfo {year} {2005})\ pp.\ \bibinfo {pages}
  {445--452}\BibitemShut {NoStop}%
\bibitem [{\citenamefont {Cai}\ \emph {et~al.}(2005{\natexlab{b}})\citenamefont
  {Cai}, \citenamefont {Shao}, \citenamefont {He}, \citenamefont {Yan},\ and\
  \citenamefont {Han}}]{cai2005mining}%
  \BibitemOpen
  \bibfield  {author} {\bibinfo {author} {\bibfnamefont {D.}~\bibnamefont
  {Cai}}, \bibinfo {author} {\bibfnamefont {Z.}~\bibnamefont {Shao}}, \bibinfo
  {author} {\bibfnamefont {X.}~\bibnamefont {He}}, \bibinfo {author}
  {\bibfnamefont {X.}~\bibnamefont {Yan}}, \ and\ \bibinfo {author}
  {\bibfnamefont {J.}~\bibnamefont {Han}},\ }in\ \href@noop {} {\emph {\bibinfo
  {booktitle} {Proceedings of the 3rd international workshop on Link
  discovery}}}\ (\bibinfo {organization} {ACM},\ \bibinfo {year} {2005})\ pp.\
  \bibinfo {pages} {58--65}\BibitemShut {NoStop}%
\bibitem [{\citenamefont {Kivel\"a}\ \emph {et~al.}(2014)\citenamefont
  {Kivel\"a}, \citenamefont {Arenas}, \citenamefont {Barthelemy}, \citenamefont
  {Gleeson}, \citenamefont {Moreno},\ and\ \citenamefont
  {Porter}}]{kivela2014}%
  \BibitemOpen
  \bibfield  {author} {\bibinfo {author} {\bibfnamefont {M.}~\bibnamefont
  {Kivel\"a}}, \bibinfo {author} {\bibfnamefont {A.}~\bibnamefont {Arenas}},
  \bibinfo {author} {\bibfnamefont {M.}~\bibnamefont {Barthelemy}}, \bibinfo
  {author} {\bibfnamefont {J.~P.}\ \bibnamefont {Gleeson}}, \bibinfo {author}
  {\bibfnamefont {Y.}~\bibnamefont {Moreno}}, \ and\ \bibinfo {author}
  {\bibfnamefont {M.~A.}\ \bibnamefont {Porter}},\ }\href {\doibase
  10.1093/comnet/cnu016} {\bibfield  {journal} {\bibinfo  {journal} {Journal of
  Complex Networks}\ }\textbf {\bibinfo {volume} {2}},\ \bibinfo {pages} {203}
  (\bibinfo {year} {2014})}\BibitemShut {NoStop}%
\bibitem [{\citenamefont {Boccaletti}\ \emph {et~al.}(2014)\citenamefont
  {Boccaletti}, \citenamefont {Bianconi}, \citenamefont {Criado}, \citenamefont
  {del Genio}, \citenamefont {G\'{o}mez-Garde\~{n}es}, \citenamefont {Romance},
  \citenamefont {Sendi\~{n}a Nadal}, \citenamefont {Wang},\ and\ \citenamefont
  {Zanin}}]{boccaletti2014}%
  \BibitemOpen
  \bibfield  {author} {\bibinfo {author} {\bibfnamefont {S.}~\bibnamefont
  {Boccaletti}}, \bibinfo {author} {\bibfnamefont {G.}~\bibnamefont
  {Bianconi}}, \bibinfo {author} {\bibfnamefont {R.}~\bibnamefont {Criado}},
  \bibinfo {author} {\bibfnamefont {C.}~\bibnamefont {del Genio}}, \bibinfo
  {author} {\bibfnamefont {J.}~\bibnamefont {G\'{o}mez-Garde\~{n}es}}, \bibinfo
  {author} {\bibfnamefont {M.}~\bibnamefont {Romance}}, \bibinfo {author}
  {\bibfnamefont {I.}~\bibnamefont {Sendi\~{n}a Nadal}}, \bibinfo {author}
  {\bibfnamefont {Z.}~\bibnamefont {Wang}}, \ and\ \bibinfo {author}
  {\bibfnamefont {M.}~\bibnamefont {Zanin}},\ }\href {\doibase
  http://dx.doi.org/10.1016/j.physrep.2014.07.001} {\bibfield  {journal}
  {\bibinfo  {journal} {Physics Reports}\ }\textbf {\bibinfo {volume} {544}},\
  \bibinfo {pages} {1} (\bibinfo {year} {2014})}\BibitemShut {NoStop}%
\bibitem [{\citenamefont {Nicosia}\ \emph {et~al.}(2013)\citenamefont
  {Nicosia}, \citenamefont {Bianconi}, \citenamefont {Latora},\ and\
  \citenamefont {Barthelemy}}]{nicosia2013}%
  \BibitemOpen
  \bibfield  {author} {\bibinfo {author} {\bibfnamefont {V.}~\bibnamefont
  {Nicosia}}, \bibinfo {author} {\bibfnamefont {G.}~\bibnamefont {Bianconi}},
  \bibinfo {author} {\bibfnamefont {V.}~\bibnamefont {Latora}}, \ and\ \bibinfo
  {author} {\bibfnamefont {M.}~\bibnamefont {Barthelemy}},\ }\href {\doibase
  10.1103/PhysRevLett.111.058701} {\bibfield  {journal} {\bibinfo  {journal}
  {Phys. Rev. Lett.}\ }\textbf {\bibinfo {volume} {111}},\ \bibinfo {pages}
  {058701} (\bibinfo {year} {2013})}\BibitemShut {NoStop}%
\bibitem [{\citenamefont {Kim}\ and\ \citenamefont {Goh}(2013)}]{kim2013}%
  \BibitemOpen
  \bibfield  {author} {\bibinfo {author} {\bibfnamefont {J.~Y.}\ \bibnamefont
  {Kim}}\ and\ \bibinfo {author} {\bibfnamefont {K.-I.}\ \bibnamefont {Goh}},\
  }\href {\doibase 10.1103/PhysRevLett.111.058702} {\bibfield  {journal}
  {\bibinfo  {journal} {Phys. Rev. Lett.}\ }\textbf {\bibinfo {volume} {111}},\
  \bibinfo {pages} {058702} (\bibinfo {year} {2013})}\BibitemShut {NoStop}%
\bibitem [{\citenamefont {Nicosia}\ \emph {et~al.}(2014)\citenamefont
  {Nicosia}, \citenamefont {Bianconi}, \citenamefont {Latora},\ and\
  \citenamefont {Barthelemy}}]{nicosia2014}%
  \BibitemOpen
  \bibfield  {author} {\bibinfo {author} {\bibfnamefont {V.}~\bibnamefont
  {Nicosia}}, \bibinfo {author} {\bibfnamefont {G.}~\bibnamefont {Bianconi}},
  \bibinfo {author} {\bibfnamefont {V.}~\bibnamefont {Latora}}, \ and\ \bibinfo
  {author} {\bibfnamefont {M.}~\bibnamefont {Barthelemy}},\ }\href {\doibase
  10.1103/PhysRevE.90.042807} {\bibfield  {journal} {\bibinfo  {journal} {Phys.
  Rev. E}\ }\textbf {\bibinfo {volume} {90}},\ \bibinfo {pages} {042807}
  (\bibinfo {year} {2014})}\BibitemShut {NoStop}%
\bibitem [{\citenamefont {Battiston}\ \emph {et~al.}(2014)\citenamefont
  {Battiston}, \citenamefont {Nicosia},\ and\ \citenamefont
  {Latora}}]{battiston2014}%
  \BibitemOpen
  \bibfield  {author} {\bibinfo {author} {\bibfnamefont {F.}~\bibnamefont
  {Battiston}}, \bibinfo {author} {\bibfnamefont {V.}~\bibnamefont {Nicosia}},
  \ and\ \bibinfo {author} {\bibfnamefont {V.}~\bibnamefont {Latora}},\ }\href
  {\doibase 10.1103/PhysRevE.89.032804} {\bibfield  {journal} {\bibinfo
  {journal} {Phys. Rev. E}\ }\textbf {\bibinfo {volume} {89}},\ \bibinfo
  {pages} {032804} (\bibinfo {year} {2014})}\BibitemShut {NoStop}%
\bibitem [{\citenamefont {Min}\ \emph {et~al.}(2014)\citenamefont {Min},
  \citenamefont {Yi}, \citenamefont {Lee},\ and\ \citenamefont
  {Goh}}]{min2014}%
  \BibitemOpen
  \bibfield  {author} {\bibinfo {author} {\bibfnamefont {B.}~\bibnamefont
  {Min}}, \bibinfo {author} {\bibfnamefont {S.~D.}\ \bibnamefont {Yi}},
  \bibinfo {author} {\bibfnamefont {K.-M.}\ \bibnamefont {Lee}}, \ and\
  \bibinfo {author} {\bibfnamefont {K.-I.}\ \bibnamefont {Goh}},\ }\href
  {\doibase 10.1103/PhysRevE.89.042811} {\bibfield  {journal} {\bibinfo
  {journal} {Phys. Rev. E}\ }\textbf {\bibinfo {volume} {89}},\ \bibinfo
  {pages} {042811} (\bibinfo {year} {2014})}\BibitemShut {NoStop}%
\bibitem [{\citenamefont {Menichetti}\ \emph {et~al.}(2014)\citenamefont
  {Menichetti}, \citenamefont {Remondini}, \citenamefont {Panzarasa},
  \citenamefont {Mondragón},\ and\ \citenamefont {Bianconi}}]{menichetti2014}%
  \BibitemOpen
  \bibfield  {author} {\bibinfo {author} {\bibfnamefont {G.}~\bibnamefont
  {Menichetti}}, \bibinfo {author} {\bibfnamefont {D.}~\bibnamefont
  {Remondini}}, \bibinfo {author} {\bibfnamefont {P.}~\bibnamefont
  {Panzarasa}}, \bibinfo {author} {\bibfnamefont {R.~J.}\ \bibnamefont
  {Mondragón}}, \ and\ \bibinfo {author} {\bibfnamefont {G.}~\bibnamefont
  {Bianconi}},\ }\href {\doibase 10.1371/journal.pone.0097857} {\bibfield
  {journal} {\bibinfo  {journal} {PLoS ONE}\ }\textbf {\bibinfo {volume} {9}},\
  \bibinfo {pages} {e97857} (\bibinfo {year} {2014})}\BibitemShut {NoStop}%
\bibitem [{\citenamefont {Nicosia}\ and\ \citenamefont
  {Latora}(2014)}]{nicosia2014measuring}%
  \BibitemOpen
  \bibfield  {author} {\bibinfo {author} {\bibfnamefont {V.}~\bibnamefont
  {Nicosia}}\ and\ \bibinfo {author} {\bibfnamefont {V.}~\bibnamefont
  {Latora}},\ }\href@noop {} {\bibfield  {journal} {\bibinfo  {journal} {arXiv
  preprint arXiv:1403.1546}\ } (\bibinfo {year} {2014})}\BibitemShut {NoStop}%
\bibitem [{\citenamefont {Norris}(1998)}]{norris1998markov}%
  \BibitemOpen
  \bibfield  {author} {\bibinfo {author} {\bibfnamefont {J.~R.}\ \bibnamefont
  {Norris}},\ }\href@noop {} {\emph {\bibinfo {title} {Markov chains}}}\
  (\bibinfo  {publisher} {Cambridge university press},\ \bibinfo {year}
  {1998})\BibitemShut {NoStop}%
\bibitem [{\citenamefont {Cellai}\ \emph {et~al.}(2013)\citenamefont {Cellai},
  \citenamefont {L\'opez}, \citenamefont {Zhou}, \citenamefont {Gleeson},\ and\
  \citenamefont {Bianconi}}]{cellai2013}%
  \BibitemOpen
  \bibfield  {author} {\bibinfo {author} {\bibfnamefont {D.}~\bibnamefont
  {Cellai}}, \bibinfo {author} {\bibfnamefont {E.}~\bibnamefont {L\'opez}},
  \bibinfo {author} {\bibfnamefont {J.}~\bibnamefont {Zhou}}, \bibinfo {author}
  {\bibfnamefont {J.~P.}\ \bibnamefont {Gleeson}}, \ and\ \bibinfo {author}
  {\bibfnamefont {G.}~\bibnamefont {Bianconi}},\ }\href {\doibase
  10.1103/PhysRevE.88.052811} {\bibfield  {journal} {\bibinfo  {journal} {Phys.
  Rev. E}\ }\textbf {\bibinfo {volume} {88}},\ \bibinfo {pages} {052811}
  (\bibinfo {year} {2013})}\BibitemShut {NoStop}%
\bibitem [{\citenamefont {Li}\ \emph {et~al.}(2013)\citenamefont {Li},
  \citenamefont {Liu}, \citenamefont {Jia},\ and\ \citenamefont
  {Wang}}]{li2013}%
  \BibitemOpen
  \bibfield  {author} {\bibinfo {author} {\bibfnamefont {M.}~\bibnamefont
  {Li}}, \bibinfo {author} {\bibfnamefont {R.-R.}\ \bibnamefont {Liu}},
  \bibinfo {author} {\bibfnamefont {C.-X.}\ \bibnamefont {Jia}}, \ and\
  \bibinfo {author} {\bibfnamefont {B.-H.}\ \bibnamefont {Wang}},\ }\href
  {http://stacks.iop.org/1367-2630/15/i=9/a=093013} {\bibfield  {journal}
  {\bibinfo  {journal} {New Journal of Physics}\ }\textbf {\bibinfo {volume}
  {15}},\ \bibinfo {pages} {093013} (\bibinfo {year} {2013})}\BibitemShut
  {NoStop}%
\bibitem [{mmc()}]{mmc-code}%
  \BibitemOpen
  \href@noop {} {}\bibinfo {howpublished}
  {\url{https://github.com/vkrmsv/MultiplexMarkovChain}}\BibitemShut {NoStop}%
\bibitem [{\citenamefont {Mansfield}\ and\ \citenamefont
  {Bronson}(1997)}]{manbro97}%
  \BibitemOpen
  \bibfield  {author} {\bibinfo {author} {\bibfnamefont {E.~D.}\ \bibnamefont
  {Mansfield}}\ and\ \bibinfo {author} {\bibfnamefont {R.}~\bibnamefont
  {Bronson}},\ }\href@noop {} {\bibfield  {journal} {\bibinfo  {journal}
  {American Political Science Review}\ }\textbf {\bibinfo {volume} {91}},\
  \bibinfo {pages} {94} (\bibinfo {year} {1997})}\BibitemShut {NoStop}%
\bibitem [{\citenamefont {Gowa}(1995)}]{gow95}%
  \BibitemOpen
  \bibfield  {author} {\bibinfo {author} {\bibfnamefont {J.}~\bibnamefont
  {Gowa}},\ }\href@noop {} {\emph {\bibinfo {title} {Allies, adversaries, and
  international trade}}}\ (\bibinfo  {publisher} {Princeton University Press},\
  \bibinfo {address} {Princeton, NJ},\ \bibinfo {year} {1995})\BibitemShut
  {NoStop}%
\bibitem [{\citenamefont {Gowa}\ and\ \citenamefont
  {Mansfield}(1993)}]{gowman93}%
  \BibitemOpen
  \bibfield  {author} {\bibinfo {author} {\bibfnamefont {J.}~\bibnamefont
  {Gowa}}\ and\ \bibinfo {author} {\bibfnamefont {E.~D.}\ \bibnamefont
  {Mansfield}},\ }\href@noop {} {\bibfield  {journal} {\bibinfo  {journal}
  {American Political Science Review}\ }\textbf {\bibinfo {volume} {87}},\
  \bibinfo {pages} {408} (\bibinfo {year} {1993})}\BibitemShut {NoStop}%
\bibitem [{\citenamefont {Gowa}\ and\ \citenamefont
  {Mansfield}(2004)}]{gowman04}%
  \BibitemOpen
  \bibfield  {author} {\bibinfo {author} {\bibfnamefont {J.}~\bibnamefont
  {Gowa}}\ and\ \bibinfo {author} {\bibfnamefont {E.~D.}\ \bibnamefont
  {Mansfield}},\ }\href@noop {} {\bibfield  {journal} {\bibinfo  {journal}
  {International Organization}\ }\textbf {\bibinfo {volume} {58}},\ \bibinfo
  {pages} {775} (\bibinfo {year} {2004})}\BibitemShut {NoStop}%
\bibitem [{\citenamefont {Long}(2003)}]{lon03}%
  \BibitemOpen
  \bibfield  {author} {\bibinfo {author} {\bibfnamefont {A.~G.}\ \bibnamefont
  {Long}},\ }\href@noop {} {\bibfield  {journal} {\bibinfo  {journal} {Journal
  of Peace Research}\ }\textbf {\bibinfo {volume} {40}},\ \bibinfo {pages}
  {537} (\bibinfo {year} {2003})}\BibitemShut {NoStop}%
\bibitem [{\citenamefont {Long}\ and\ \citenamefont {Leeds}(2006)}]{lonlee06}%
  \BibitemOpen
  \bibfield  {author} {\bibinfo {author} {\bibfnamefont {A.~G.}\ \bibnamefont
  {Long}}\ and\ \bibinfo {author} {\bibfnamefont {B.~A.}\ \bibnamefont
  {Leeds}},\ }\href@noop {} {\bibfield  {journal} {\bibinfo  {journal} {Journal
  of Peace Research}\ }\textbf {\bibinfo {volume} {43}},\ \bibinfo {pages}
  {433} (\bibinfo {year} {2006})}\BibitemShut {NoStop}%
\bibitem [{\citenamefont {Morrow}\ \emph {et~al.}(1998)\citenamefont {Morrow},
  \citenamefont {Siverson},\ and\ \citenamefont {Tabares}}]{morsiv98}%
  \BibitemOpen
  \bibfield  {author} {\bibinfo {author} {\bibfnamefont {J.~D.}\ \bibnamefont
  {Morrow}}, \bibinfo {author} {\bibfnamefont {R.~M.}\ \bibnamefont
  {Siverson}}, \ and\ \bibinfo {author} {\bibfnamefont {T.}~\bibnamefont
  {Tabares}},\ }\href@noop {} {\bibfield  {journal} {\bibinfo  {journal}
  {American Political Science Review}\ }\textbf {\bibinfo {volume} {92}},\
  \bibinfo {pages} {649} (\bibinfo {year} {1998})}\BibitemShut {NoStop}%
\bibitem [{\citenamefont {Fordham}(2010)}]{for10}%
  \BibitemOpen
  \bibfield  {author} {\bibinfo {author} {\bibfnamefont {B.~O.}\ \bibnamefont
  {Fordham}},\ }\href@noop {} {\bibfield  {journal} {\bibinfo  {journal}
  {Journal of Peace Research}\ }\textbf {\bibinfo {volume} {47}},\ \bibinfo
  {pages} {685} (\bibinfo {year} {2010})}\BibitemShut {NoStop}%
\bibitem [{\citenamefont {Maoz}(2010)}]{maoz2010}%
  \BibitemOpen
  \bibfield  {author} {\bibinfo {author} {\bibfnamefont {Z.}~\bibnamefont
  {Maoz}},\ }\href@noop {} {\emph {\bibinfo {title} {Networks of nations: The
  evolution, structure, and impact of international networks, 1816--2001}}},\
  Vol.~\bibinfo {volume} {32}\ (\bibinfo  {publisher} {Cambridge University
  Press},\ \bibinfo {year} {2010})\BibitemShut {NoStop}%
\bibitem [{\citenamefont {Gharehyazie}\ \emph {et~al.}(2014)\citenamefont
  {Gharehyazie}, \citenamefont {Posnett}, \citenamefont {Vasilescu},\ and\
  \citenamefont {Filkov}}]{gharehyazie2014}%
  \BibitemOpen
  \bibfield  {author} {\bibinfo {author} {\bibfnamefont {M.}~\bibnamefont
  {Gharehyazie}}, \bibinfo {author} {\bibfnamefont {D.}~\bibnamefont
  {Posnett}}, \bibinfo {author} {\bibfnamefont {B.}~\bibnamefont {Vasilescu}},
  \ and\ \bibinfo {author} {\bibfnamefont {V.}~\bibnamefont {Filkov}},\
  }\href@noop {} {\bibfield  {journal} {\bibinfo  {journal} {Empirical Software
  Engineering}\ ,\ \bibinfo {pages} {1}} (\bibinfo {year} {2014})}\BibitemShut
  {NoStop}%
\bibitem [{\citenamefont {Nakakoji}\ \emph {et~al.}(2002)\citenamefont
  {Nakakoji}, \citenamefont {Yamamoto}, \citenamefont {Nishinaka},
  \citenamefont {Kishida},\ and\ \citenamefont {Ye}}]{nakakoji2002}%
  \BibitemOpen
  \bibfield  {author} {\bibinfo {author} {\bibfnamefont {K.}~\bibnamefont
  {Nakakoji}}, \bibinfo {author} {\bibfnamefont {Y.}~\bibnamefont {Yamamoto}},
  \bibinfo {author} {\bibfnamefont {Y.}~\bibnamefont {Nishinaka}}, \bibinfo
  {author} {\bibfnamefont {K.}~\bibnamefont {Kishida}}, \ and\ \bibinfo
  {author} {\bibfnamefont {Y.}~\bibnamefont {Ye}},\ }in\ \href@noop {} {\emph
  {\bibinfo {booktitle} {Proceedings of the international workshop on
  Principles of software evolution}}}\ (\bibinfo {organization} {ACM},\
  \bibinfo {year} {2002})\ pp.\ \bibinfo {pages} {76--85}\BibitemShut {NoStop}%
\bibitem [{\citenamefont {Ducheneaut}(2005)}]{ducheneaut2005}%
  \BibitemOpen
  \bibfield  {author} {\bibinfo {author} {\bibfnamefont {N.}~\bibnamefont
  {Ducheneaut}},\ }\href@noop {} {\bibfield  {journal} {\bibinfo  {journal}
  {Computer Supported Cooperative Work (CSCW)}\ }\textbf {\bibinfo {volume}
  {14}},\ \bibinfo {pages} {323} (\bibinfo {year} {2005})}\BibitemShut
  {NoStop}%
\bibitem [{\citenamefont {Xuan}\ \emph {et~al.}(2012)\citenamefont {Xuan},
  \citenamefont {Gharehyazie}, \citenamefont {Devanbu},\ and\ \citenamefont
  {Filkov}}]{xuan2012}%
  \BibitemOpen
  \bibfield  {author} {\bibinfo {author} {\bibfnamefont {Q.}~\bibnamefont
  {Xuan}}, \bibinfo {author} {\bibfnamefont {M.}~\bibnamefont {Gharehyazie}},
  \bibinfo {author} {\bibfnamefont {P.}~\bibnamefont {Devanbu}}, \ and\
  \bibinfo {author} {\bibfnamefont {V.}~\bibnamefont {Filkov}},\ }in\ \href
  {\doibase 10.1109/SocialInformatics.2012.17} {\emph {\bibinfo {booktitle}
  {Social Informatics (SocialInformatics), 2012 International Conference on}}}\
  (\bibinfo {year} {2012})\ pp.\ \bibinfo {pages} {78--85}\BibitemShut
  {NoStop}%
\bibitem [{\citenamefont {Xuan}\ and\ \citenamefont {Filkov}(2014)}]{xuan2014}%
  \BibitemOpen
  \bibfield  {author} {\bibinfo {author} {\bibfnamefont {Q.}~\bibnamefont
  {Xuan}}\ and\ \bibinfo {author} {\bibfnamefont {V.}~\bibnamefont {Filkov}},\
  }in\ \href {\doibase 10.1145/2568225.2568238} {\emph {\bibinfo {booktitle}
  {Proceedings of the 36th International Conference on Software
  Engineering}}},\ \bibinfo {series and number} {ICSE 2014}\ (\bibinfo
  {publisher} {ACM},\ \bibinfo {address} {New York, NY, USA},\ \bibinfo {year}
  {2014})\ pp.\ \bibinfo {pages} {222--233}\BibitemShut {NoStop}%
\bibitem [{apa()}]{apache}%
  \BibitemOpen
  \href@noop {} {}\bibinfo {howpublished}
  {\url{(http://www.apache.org/)}}\BibitemShut {NoStop}%
\bibitem [{\citenamefont {Sivia}\ and\ \citenamefont
  {Skilling}(2006)}]{sivia2006data}%
  \BibitemOpen
  \bibfield  {author} {\bibinfo {author} {\bibfnamefont {D.}~\bibnamefont
  {Sivia}}\ and\ \bibinfo {author} {\bibfnamefont {J.}~\bibnamefont
  {Skilling}},\ }\href@noop {} {\emph {\bibinfo {title} {Data Analysis: A
  Bayesian Tutorial}}}\ (\bibinfo  {publisher} {Oxford University Press, USA},\
  \bibinfo {year} {2006})\BibitemShut {NoStop}%
\bibitem [{\citenamefont {Leeds}(2005)}]{leeds2005}%
  \BibitemOpen
  \bibfield  {author} {\bibinfo {author} {\bibfnamefont {B.~A.}\ \bibnamefont
  {Leeds}},\ }\href@noop {} {\bibfield  {journal} {\bibinfo  {journal} {Rice
  University, Department of Political Science, Houston}\ } (\bibinfo {year}
  {2005})}\BibitemShut {NoStop}%
\bibitem [{\citenamefont {Barbieri}\ \emph {et~al.}(2009)\citenamefont
  {Barbieri}, \citenamefont {Keshk},\ and\ \citenamefont
  {Pollins}}]{barbieri2009}%
  \BibitemOpen
  \bibfield  {author} {\bibinfo {author} {\bibfnamefont {K.}~\bibnamefont
  {Barbieri}}, \bibinfo {author} {\bibfnamefont {O.~M.}\ \bibnamefont {Keshk}},
  \ and\ \bibinfo {author} {\bibfnamefont {B.~M.}\ \bibnamefont {Pollins}},\
  }\href@noop {} {\bibfield  {journal} {\bibinfo  {journal} {Conflict
  Management and Peace Science}\ }\textbf {\bibinfo {volume} {26}},\ \bibinfo
  {pages} {471} (\bibinfo {year} {2009})}\BibitemShut {NoStop}%
\bibitem [{\citenamefont {Gleditsch}(2002)}]{gleditsch2002}%
  \BibitemOpen
  \bibfield  {author} {\bibinfo {author} {\bibfnamefont {K.~S.}\ \bibnamefont
  {Gleditsch}},\ }\href@noop {} {\bibfield  {journal} {\bibinfo  {journal}
  {Journal of Conflict Resolution}\ }\textbf {\bibinfo {volume} {46}},\
  \bibinfo {pages} {712} (\bibinfo {year} {2002})}\BibitemShut {NoStop}%
\bibitem [{oss()}]{oss-email-url}%
  \BibitemOpen
  \href@noop {} {}\bibinfo {howpublished}
  {\url{http://mail-archives.apache.org/mod_mbox/}}\BibitemShut {NoStop}%
\end{thebibliography}
\end{document}